\documentclass{solarphysics_edit}
\usepackage[nointegrals]{wasysym}
\usepackage{amsmath}
\usepackage[hyperref,optionalrh,solaromanenum]{spr-sola-addons} 
\usepackage{graphicx}                    
\usepackage[normalem]{ulem}
\usepackage{rotating}

\usepackage{units}
\usepackage{lmodern}
\usepackage{comment}
\wasyfamily
\DeclareFontShape{U}{wasy}{b}{n}{ <-10> ssub * wasy/m/n
	<10> <10.95> <12> <14.4> <17.28> <20.74> <24.88>wasyb10 }{}
\usepackage{xcolor}
\usepackage{hyperref}
\hypersetup{%
	colorlinks,
	linkcolor={red!50!black},
	citecolor={blue!50!black},
	urlcolor={blue!80!black},
	pdfborder={0 0 0},
	urlcolor=blue,
	breaklinks=true
}

\newcommand{\D}{\mathrm{d}}

\begin{document}

\begin{article}

\begin{opening}

\title{GONG p-mode parameters through two solar cycles}

%
\author[addressref={aff1,aff2,aff3},corref,email={R.Kiefer@warwick.ac.uk}]{\inits{R.}\fnm{Ren\'e}~\lnm{Kiefer}\orcid{0000-0003-4166-5343}}
\author[addressref=aff3, corref, email={rkomm@nso.edu}, ]{\inits{R.}\fnm{Rudi}~\lnm{Komm}\orcid{0000-0001-5736-1868}}
\author[addressref=aff3]{\inits{F.}\fnm{Frank}~\lnm{Hill}\orcid{0000-0001-6714-8681}}
\author[addressref=aff1]{\inits{A.-M.}\fnm{Anne-Marie}~\lnm{Broomhall}\orcid{0000-0002-5209-9378}}
\author[addressref=aff2]{\inits{M.}\fnm{Markus}~\lnm{Roth}\orcid{0000-0002-1430-7172}}
%
\runningauthor{R. Kiefer {\it et al.}}
\runningtitle{GONG p-mode parameters}


\address[id=aff1]{Centre for Fusion, Space, and Astrophysics, Department of Physics, University of Warwick,	Coventry, CV4 7AL, UK}
\address[id=aff2]{Kiepenheuer-Institut f\"ur Sonnenphysik, Sch\"oneckstra\ss e 6, 79104, Freiburg, Germany}
\address[id=aff3]{National Solar Observatory, 3665 Discovery Drive, Boulder, CO 80303}

\begin{abstract}
	We investigate the parameters of global solar p-mode oscillations, namely damping width $\Gamma$, amplitude $A$, mean squared velocity $\langle v^2\rangle$, energy $E$, and energy supply rate $\D E/\D t$, derived from two solar cycles' worth (1996--2018) of Global Oscillation Network Group (GONG) time series for harmonic degrees \nobreak{$l=0$--$150$}. We correct for the effect of fill factor, apparent solar radius, and spurious jumps in the mode amplitudes. We find that the amplitude of the activity related changes of $\Gamma$ and $A$ depends on both frequency and harmonic degree of the modes, with the largest variations of $\Gamma$ for modes with $\unit[2400]{\mu Hz}\le \nu \le \unit[3300]{\mu Hz}$ and $31\le l \le 60$ with a min-to-max variation of $26.6\pm0.3\%$ and of $A$ for modes with $\unit[2400]{\mu Hz}\le \nu \le \unit[3300]{\mu Hz}$ and $61\le l \le 100$ with a min-to-max variation of $27.4\pm0.4\%$. The level of correlation between the solar radio flux $F_{10.7}$ and mode parameters also depends on mode frequency and harmonic degree. As a function of mode frequency, the mode amplitudes are found to follow an asymmetric Voigt profile with $\nu_{\text{max}}=\unit[3073.59\pm0.18]{\mu Hz}$. From the mode parameters, we calculate physical mode quantities and average them over specific mode frequency ranges. This way, we find that the mean squared velocities $\langle v^2\rangle$ and energies $E$ of p modes are anti-correlated with the level of activity, varying by $14.7\pm0.3\%$ and $18.4\pm0.3\%$, respectively, and that the mode energy supply rates show no significant correlation with activity. With this study we expand previously published results on the temporal variation of solar p-mode parameters. Our results will be helpful to future studies of the excitation and damping of p modes, i.e., the interplay between convection, magnetic field, and resonant acoustic oscillations. 
\end{abstract}

%

\end{opening}

%
 \section{Introduction}\label{sec:1} 
The properties of near-surface convection are subject to slight changes over the course of the activity cycle (e.g., a decrease in granule size in phase with the solar cycle found by \citealp{Macris1984,Muller1988}, and a decrease of granular contrast with increasing level of magnetic activity, see \citealp{Muller2007}). As solar p modes are stochastically driven by the acoustic noise generated by convective motion, also the related p-mode parameters vary over the solar activity cycle. However, most previous studies that investigated these activity related changes of solar p-mode parameters, aside from mode frequencies, were either limited to unresolved solar time series and were thus restricted to low harmonic degrees $l\lesssim3$ \citep[e.g.,][]{Palle1990, Elsworth1993, Chaplin1998a, Howe2003, JimenezReyes2004, Salabert2007, Broomhall2015}, or had a very short data baseline to work on (e.g., \citealp{Jefferies1991} with \unit[460]{hr} of data with a 54\% duty cycle). Over the last years, little attention has been given to the activity related changes of solar p-mode parameters for resolved data aside from mode frequencies and frequency splitting coefficients \citep[the exeptions being][]{Komm2000, Korzennik2013, Korzennik2017}. Now, with two full solar sunspot cycles' worth of data from the Global Oscillation Network Group (GONG) available, we examine the temporal variations of parameters of solar p modes of harmonic degrees $l= 0$--$150$. 

It is well established that mode widths, $\Gamma$, which are inversely proportional to the lifetimes of the mode, vary in phase with the level of solar magnetic activity \citep[e.g.,][]{Jefferies1991, Komm2000a, Chaplin2000, Jimenez2002, JimenezReyes2003, JimenezReyes2004, Salabert2007, Burtseva2009a, Broomhall2015}. This is taken to indicate that the oscillations are damped by the presence of magnetic field. The mode amplitudes, $A$, are observed to be in anti-phase with magnetic activity \citep[e.g.,][]{Palle1990, Elsworth1993,Chaplin2000, Komm2000a, Jimenez2002, JimenezReyes2003, JimenezReyes2004, Broomhall2015}. The magnitude of the change of $\Gamma$ and $A$ depends on both mode frequency and harmonic degree. Along with mode amplitudes, the mean squared velocity $\langle v^2\rangle$ and energy $E$ of the p modes change with the solar cycle, where highest mode velocities and energies are observed during solar minimum \citep{Komm2000, JimenezReyes2003,Salabert2007, JimenezReyes2004}. The energy supply rate to the modes, which is proportional to mode energy times mode width, has been measured not to change through the solar cycle for modes of low harmonic degree \citep{Chaplin2000, JimenezReyes2003, JimenezReyes2004, Broomhall2015}. 

In this article, we do not consider mode frequencies. As these are more accessible than the parameters studied here, they have been subject of numerous studies in the past and their behaviour along the solar activity cycle is well documented for a wide range of harmonic degrees \citep[e.g.,][]{Woodard1985, Elsworth1990, Libbrecht1990, Jimenez-Reyes1998, Chaplin2001, Salabert2015a, Tripathy2015, Broomhall2017}. Mode frequencies are correlated with solar activity, being at their highest during times of strong activity. For very high frequency modes, around and above the acoustic cut-off frequency, this correlation turns into an anti-correlation: the frequencies of these modes are higher during times of weak activity \citep[see, e.g., ][]{Woodard1991, Howe2008a, Rhodes2010}. The magnitudes of the shifts of mode frequencies depend on both mode frequency and harmonic degree (see \cite{Basu2016} and references therein).

This article is structured as follows: The data and the steps to correct are described in Section~\ref{sec:2}. This includes correcting for the spatial masking of GONG Dopplergrams and azimuthal averaging (Section~\ref{sec:2:2}), and measures to account for the imperfect duty cycle (Section~\ref{sec:2:3}) and for spurious jumps in the mode parameters (Section~\ref{sec:2:4}). We present our results for the temporal variation of mode widths and amplitudes in Section~\ref{sec:3} and focus on the temporal average of the mode amplitudes in Section~\ref{sec:3:3}. We go on to consider the temporal variation of the quantities mean squared velocity, energy, and energy supply rate of the p modes (Section~\ref{sec:4}). A summary and discussion of our findings is given in Section~\ref{sec:5}.

\section{Data and Methodology}\label{sec:2}
\subsection{Data Sets}\label{sec:2:1}
	In the present analysis, we use mode parameter data which were produced by the standard GONG pipeline \citep{Anderson1990, Hill1996, Hill1998} from solar full-disk Dopplergrams\footnote{The data are publicly available online and can be downloaded here: \url{ftp://gong2.nso.edu/TSERIES/}.}. The mode parameters are obtained by fitting symmetric Lorentzian profiles to the power spectra of 108-day long data sets, where every third data set is independent, i.e., they overlap by 72 days. These time series are concatenations of three GONG months, with one GONG month being 36 days. This ensures a reasonable frequency resolution of the fitted power spectra and number of independent data points. The power spectra are computed and fitted for all harmonic degrees $l$ and azimuthal orders $-l\le m\le l$ up to $l=150$. The model that is fitted to each peak in the power spectra consists of a linear two parameter background and a Lorentzian profile, which depends on the parameters width $\Gamma$, amplitude $A$, and frequency $\nu$. Mode frequencies and the background parameters are not considered in the following. The GONG peak-fitting algorithm fits symmetric Lorentzian profiles to the spectra \citep{Hill1996}. Since the mode peaks are known to be asymmetric \citep[e.g.,][]{Nigam1998}, this might introduce small, temporally varying systematics to the mode parameter analyzed, as the asymmetry changes over the solar cycle \citep{JimenezReyes2007, Korzennik2013a, Howe2015}. 
	
	\begin{figure}
		\begin{center}
			\includegraphics[width=\textwidth]{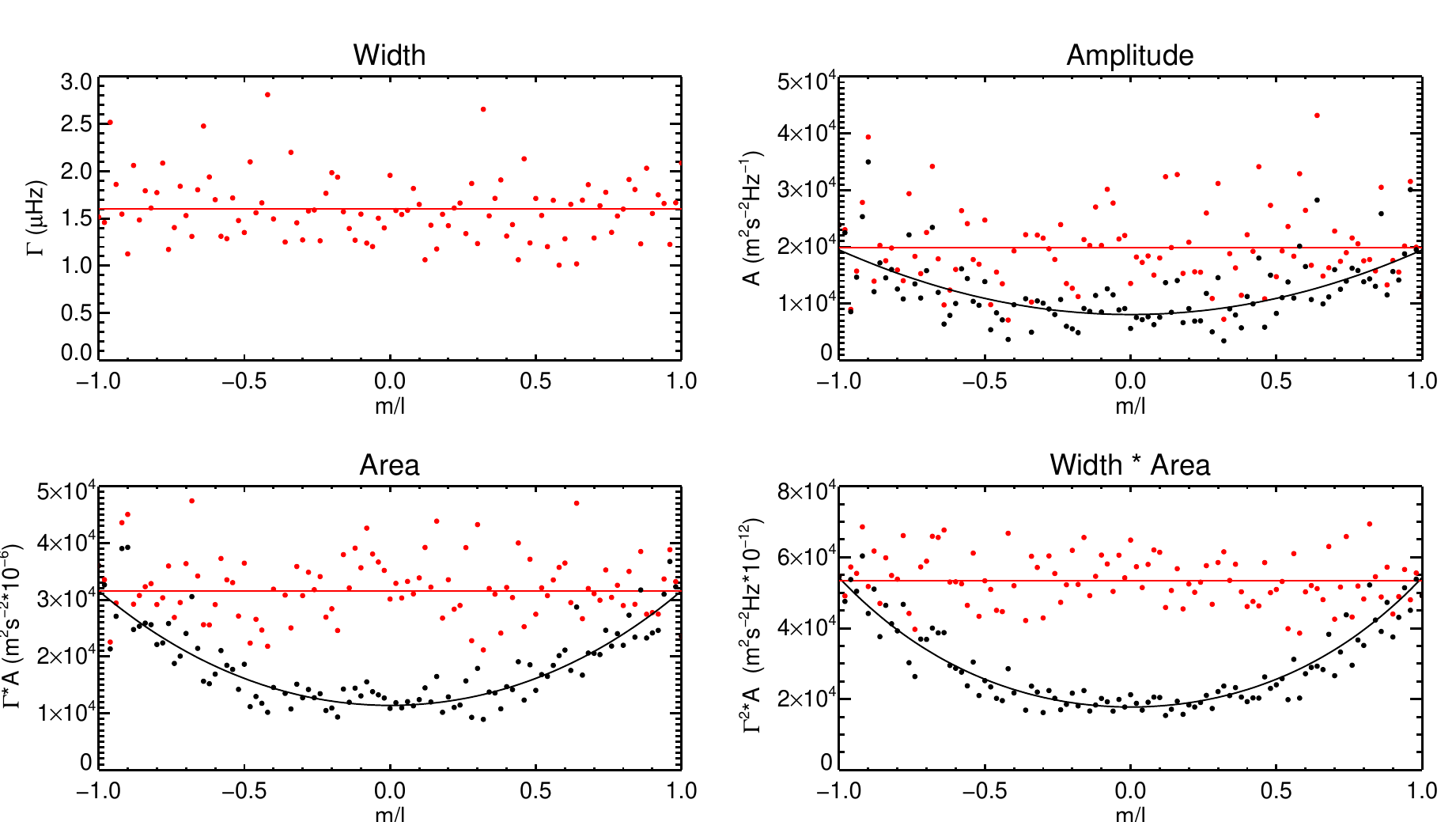}
		\end{center}
		\caption{Correction for the effects of the spatial masking for the mode $(n,l)=(10,50)$ in GONG month 5. Black data points are measurement values. Solid black lines are polynomial fits to the data to account for the effect of the spatial masking. Red data points are the corrected values. Solid red lines indicate the weighted mean of the corrected data.}
		\label{fig:1}
	\end{figure}
	
\subsection{Azimuthal Averaging and Correction for Spatial Masking}\label{sec:2:2}
	In full-disk images of the Sun, pixels close to the solar limb are subject to several unwanted effects. First, because GONG measures line-of-sight velocities, projection effects increase towards the solar limb. Second, the resolution of each pixel, measured in distance on the solar surface, increases towards the limb. A spatial mask is applied in the GONG pipeline to cut away these unwanted pixels. This leads to a suppression of mode amplitudes depending on the ratio of azimuthal order and harmonic degree $|m/l|$: While spherical harmonics with $m=0$ are oscillating over all latitudes (with a temporal dependence of $e^{-i\omega t}$, where $\omega$ is the angular frequency), spherical harmonics with azimuthal order $m\ne 0$ are confined to latitudes closer to the equator. The confinement to low latitudes becomes more pronounced for higher harmonic degree and higher $|m/l|$. Therefore, by masking out regions near the solar limb, more mode power is cut away from modes with low $|m/l|$ than from those with higher $|m/l|$. In contrast, the damping width of a mode in a power spectrum, representing the lifetime of the mode, is negligibly affected by the masking.
	
	Figure~\ref{fig:1} shows the outputs of the GONG pipeline for mode width $\Gamma$, amplitude $A$, as well as the products $\Gamma \cdot A$ and $\Gamma^2\cdot A$ of the mode $(n,l)=(10,50)$ from GONG month 5, where $n$ is the radial order. In the top left panel, the mode widths of the azimuthal components are shown as a function of $m/l$ but no visible dependence is observed. The solid red line is the error-weighted mean of the data. In the top right panel, the black data points are the measured values of the mode amplitudes. As discussed above, mode amplitudes and products which include mode amplitudes show a dependence on $m/l$, which is introduced by the spatial mask. It is accounted for by fitting a polynomial proportional to $(m/l)^{2k}$ with $k=0,1,2$ to the data. This polynomial function is empirically determined \citep[see][]{Komm2000a}. The resulting fit is shown as a solid black line in the top right and two lower panels of Figure~\ref{fig:1}. The dependence is then removed, where the values at $|m/l|=1$ are kept constant. The error-weighted mean of the corrected data points is adopted as the representative value of the mode parameter (shown as a solid red line). In order to obtain meaningful weighted mean values, in the subsequent analysis, only those multiplets are included for which at least one third of the azimuthal components are fitted by the GONG pipeline. This minimum number of fitted components is set to two thirds for modes with low harmonic degree $l\le 10$, to account for their small number of azimuthal components. In the following, we avoid the term \textit{multiplet}, as only azimuthal averages are considered, and we use the term \textit{mode} instead.
	
	\begin{figure}
		\begin{center}
			\includegraphics[width=\textwidth]{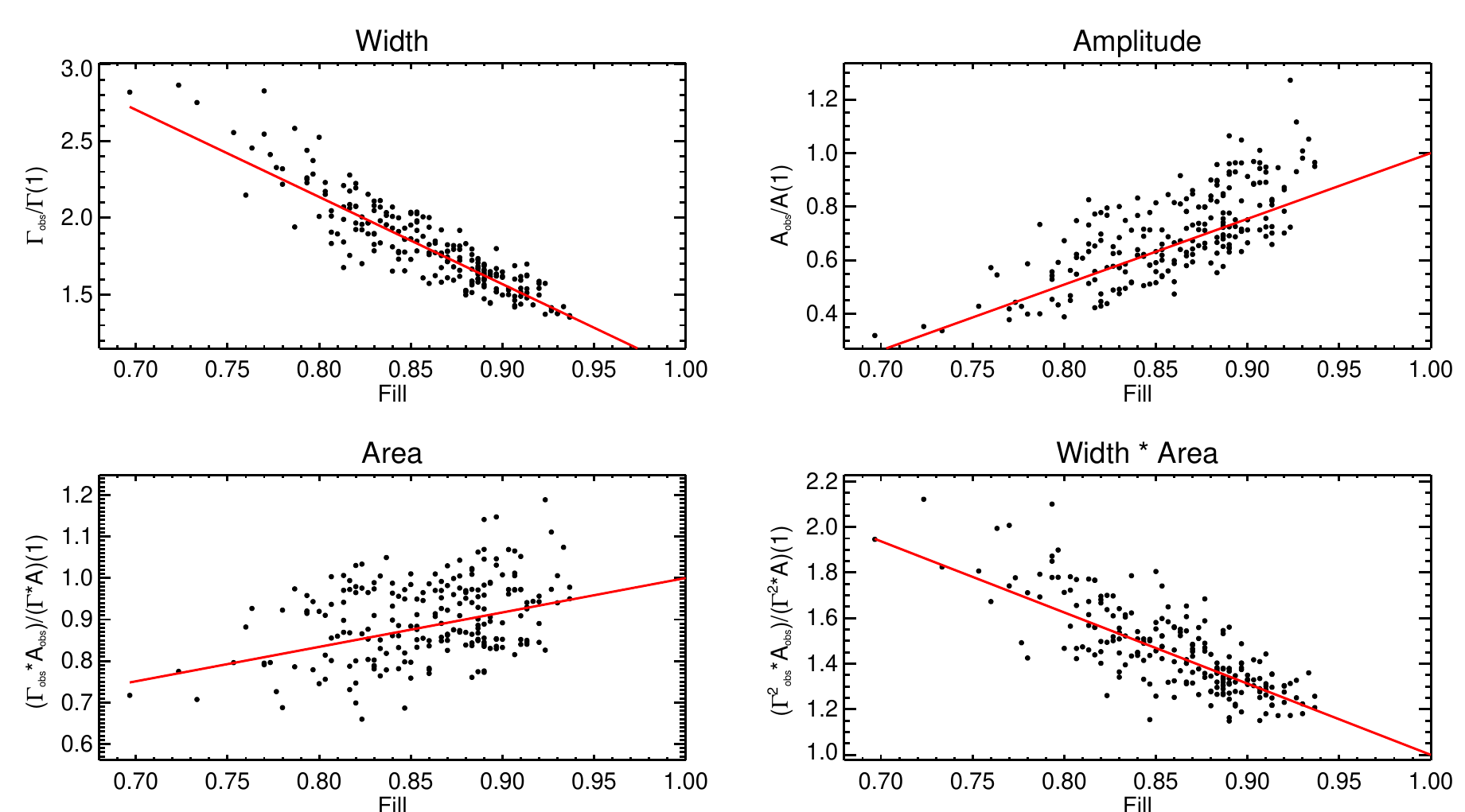}
		\end{center}
		\caption{Correction of the effect of the temporal window function on mode parameters of mode $(n,l)=(10,50)$. Parameter values are shown as a function of the temporal fill factor of the individual GONG months. Linear fits to the data are shown in red lines. Data are normalized to the value of the linear fit at fill\,=\,1. }
		\label{fig:2}
	\end{figure}
	
\subsection{Correction for Window Function}\label{sec:2:3}
	The temporal fill factor of GONG is lower than unity. This leads to a redistribution of the power from the mode peak to neighboring frequency bins and into side lobes of the main peak, which, in turn, results in a diminished mode amplitude and an increased mode width. The side lobes are caused by repeating gaps in the data. These side lobes, especially those caused by daily gaps, are considerably suppressed for a network distributed around the globe as GONG is \citep{Hill1996, Leibacher1999}. A machine-readable table with the fill factor of each 36-day GONG month is available at \url{https://gong2.nso.edu/fill.txt}. The specific structure of the gaps does not have to be taken into account as the fill factor is rather high with values between 69\% and 93\%. To account for the impact of gaps, we adopted the approach described in \citet{Komm2000}, who showed that a linear regression is sufficient to correct for the effect of the temporal window function of GONG data.  
	
	In Figure~\ref{fig:2}, the mode width $\Gamma$, mode amplitude $A$, mode area $\Gamma\cdot A$, and mode width times mode area $\Gamma^2\cdot A$ of the mode $(n,l)=(10,50)$ are shown as a function of fill factor. The products $\Gamma\cdot A$ and $\Gamma^2\cdot A$ will be utilized in Section~\ref{sec:4} to calculate the quantities of mean squared velocity, mode energy, and energy supply rate of the modes. The data are normalized to the value of the linear regression at fill\,=\,1. To obtain the corrected parameter values, the resulting fit of a linear function, shown as a red line in each panel of Figure~\ref{fig:2}, is first subtracted from the data and then its value at fill\,=\,1 is added. As previously demonstrated by, e.g., \citet{Komm2000a}, mode widths increase with decreasing fill factor and amplitudes decrease with decreasing fill factor. It should be noted that the background amplitude increases with decreasing fill factor. This indicates that power from the mode peaks is indeed distributed into the background. 

\subsection{Correction for Jumps in Sensitivity}\label{sec:2:4}
	There are two jumps in the mode amplitudes which we corrected for with an empirical correction factor. The first jump around GONG month 60 is due to a camera upgrade. The second jump around month 100 is of uncertain origin. The correction factor $C$ is given in Equation~\ref{correction:factors:amp} in Appendix~\ref{app:a}. More details about the applied corrections and the jumps are given in Appendix~\ref{app:a}.
	
\subsection{Proxy for Solar Activity}\label{sec:2:5}
	Many quantities connected to the Sun are known to vary with the solar activity cycle, e.g., the sunspot number and the emission in the Ca~\textsc{II} H \& K lines. Here, we use the solar radio flux $F_{10.7}$ as a proxy for the level of solar activity\footnote{The $F_{10.7}$ time series that was used in this work can be downloaded from this website: \url{spaceweather.gc.ca/solarflux/sx-5-en.php}}. The $F_{10.7}$ is the total emission on the solar disk at a wavelength of $\unit[10.7]{cm}$ integrated over one hour.  $F_{10.7}$ is measured in \textit{solar flux units} $\unit{sfu}$, where $\unit[1]{sfu} = \unit[10^{-22}]{W\,m^{-2}\,Hz^{-1}}$. It is a proxy for the level of activity in the upper chromosphere and the lower corona \citep{Tapping2013}. A comparison of different proxies of solar magnetic activity and their correlation with the activity related frequency shifts of solar p modes over the last 3 solar cycles can be found in \citet{Broomhall2015a}.
	
	\begin{figure}
		\begin{center}
			\includegraphics[width=\textwidth,clip=]{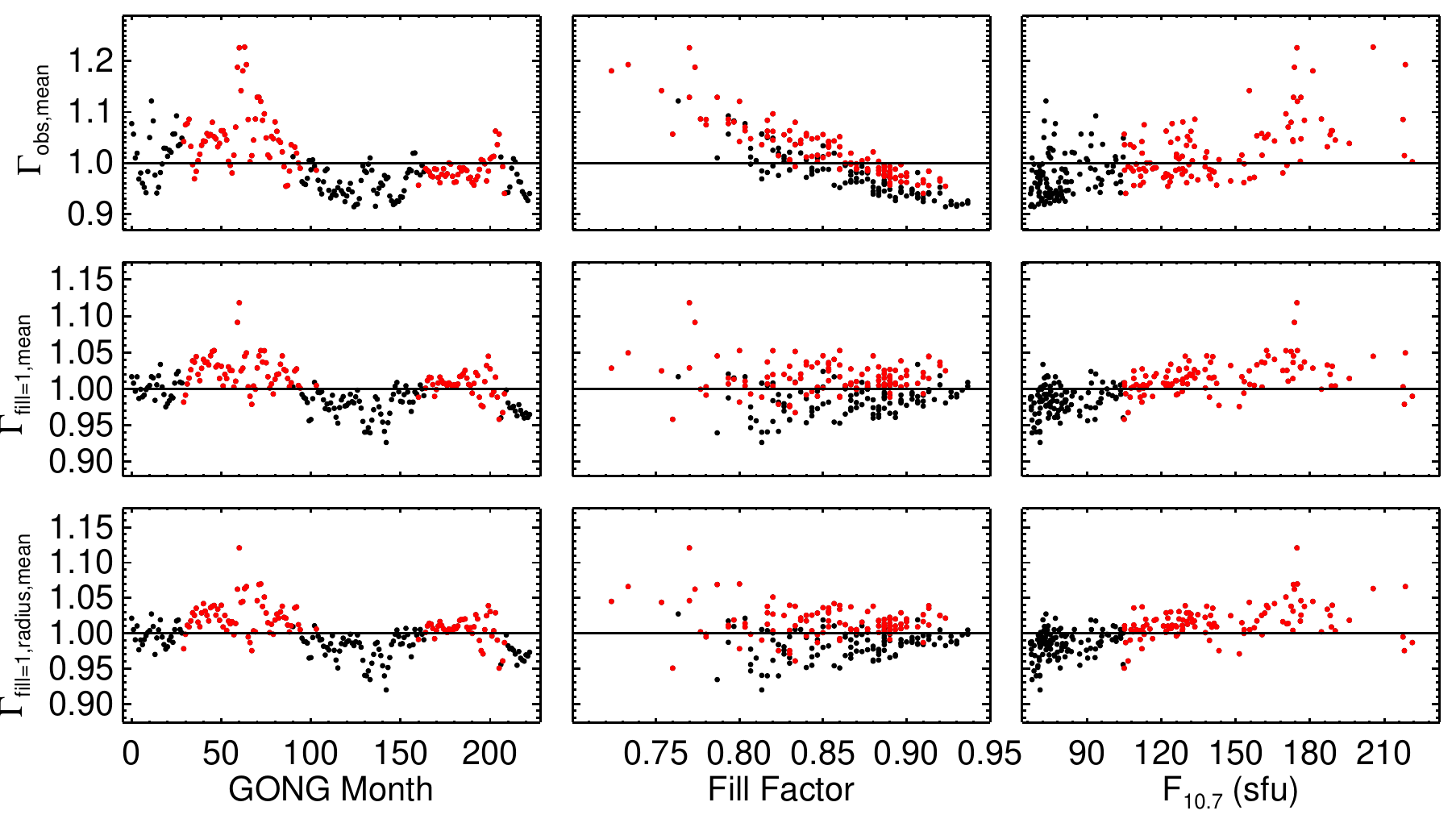}
		\end{center}
		\caption{Normalized mode widths averaged over all modes as a function of time (\textit{left column}), fill factor (\textit{middle column}), and of the magnetic activity proxy $F_{10.7}$ (\textit{right column}). The top row shows the raw data, the middle row is normalized for the effects of the fill factor, and the bottom row is corrected for fill and for change in apparent radius of the Sun.}
		\label{fig:3}
	\end{figure}
	
	\begin{figure}
		\begin{center}
			\includegraphics[width=\textwidth]{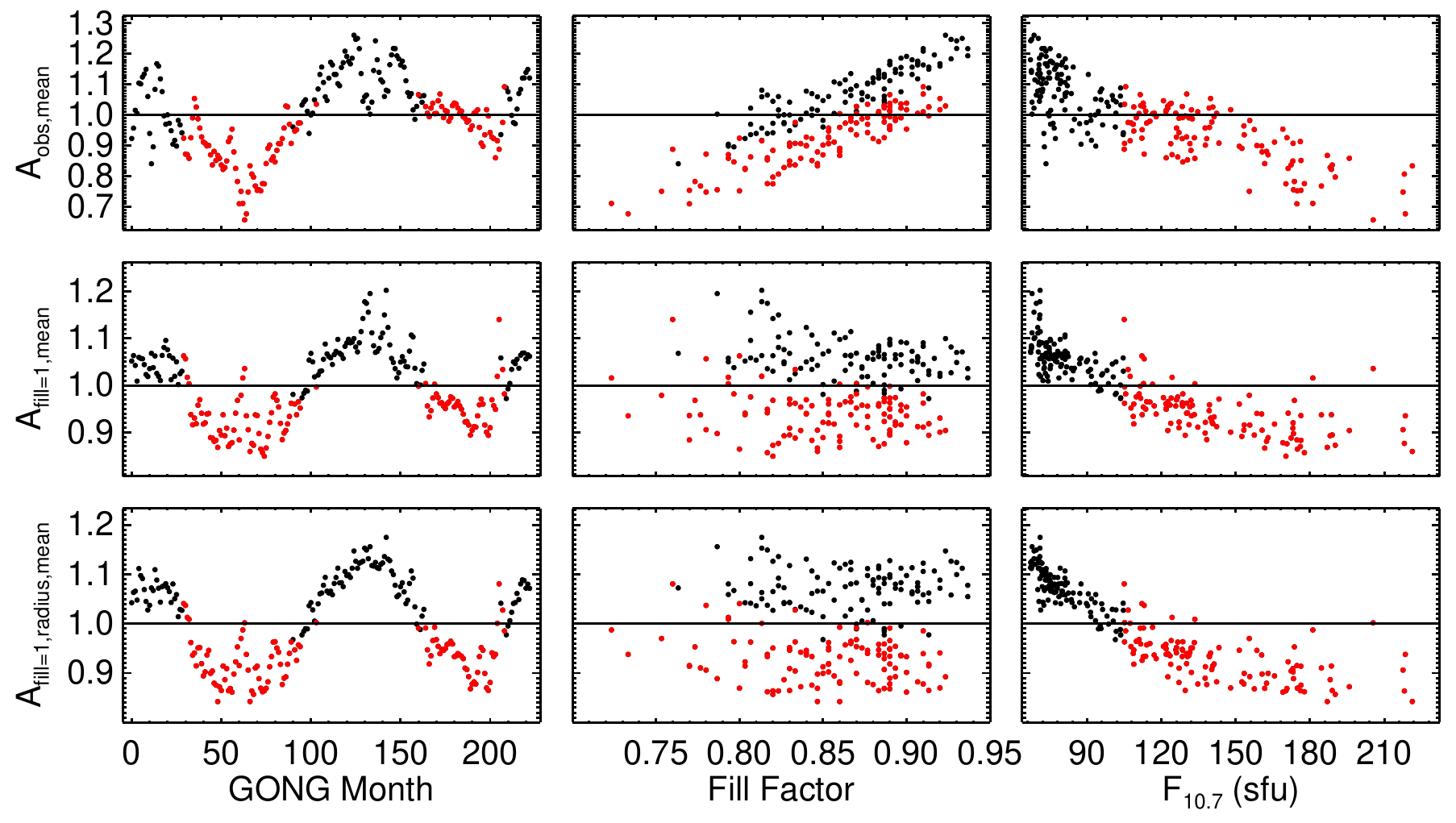}
		\end{center}
		\caption{Same as Figure~\ref{fig:3} but for mode amplitudes.}
		\label{fig:4}
	\end{figure}
	
	\section{Temporal Variation of Mode Parameters}\label{sec:3}
	\subsection{Mode Widths}\label{sec:3:1}
	To obtain clear results on their temporal variation, mode parameters have to be averaged over many modes. From here on, only modes which are present at all time samples are considered. The analysis of both mode widths and mode amplitudes is restricted to modes with frequencies $\unit[1500]{\mu Hz}\le \nu \le \unit[4500]{\mu Hz}$ and harmonic degree $0\le l \le 150$. This results in a set of 1275 mode widths and 1272 mode amplitudes. Different error flags for bad mode widths and amplitudes result in the small difference of number of modes in the sets. As both mode widths and amplitudes are dependent on mode frequency and harmonic degree, they are normalized to the mean over all time samples for each mode individually. Next, the normalized parameters of all included modes are averaged. The result for the observed mode widths $\Gamma_{\text{obs, mean}}$ is shown in the first panel of the top row of Figure~\ref{fig:3}. Here, the correction for the effect of the fill factor has not yet been applied. The need for such a correction can be seen from the middle panel of the top row, where the same data are plotted as a function of fill factor. The linear dependence on fill is apparent. In the right panel, the mode widths are shown as a function of the $F_{10.7}$ solar radio flux. The red data points in all panels of Figure~\ref{fig:3} indicate time samples with higher than median level of activity as computed from the $F_{10.7}$ flux. 
	
	In the middle row of Figure~\ref{fig:3}, the mode widths $\Gamma_{\text{fill=1, mean}}$ are shown after the correction for the temporal window function. A second correction is required to account for the apparent change in solar radius along the year. This change affects the measured values of oscillation parameters as the spatial resolution of the Dopplergrams changes with it. We apply a linear correction to the mode parameters as a function of apparent solar radius to correct for this. The bottom row shows the mode widths $\Gamma_{\text{fill=1, radius, mean}}$ after both the effect of the fill factor and the change of the apparent solar radius have been removed. 
	
	For better visibility, the left panel of the bottom row of Figure~\ref{fig:3} is shown again in the top panel of Figure~\ref{fig:5}. As can be seen, the uncertainties on the normalized variation of the mode widths are of the order of a few tenth of a percent. The error bars given here are computed as the standard error of the mean. The solid black line is the one-year running average. The red line is the $F_{10.7}$ solar radio flux. It is boxcar smoothed over one year and scaled to match the extrema of the one-year boxcar smoothed variation of the mode widths. The correlation coefficient (Spearman's rank correlation coefficient $\rho$) between the variation of the width and the $F_{10.7}$ index (both unsmoothed) is $\rho=0.62$ with $p= 2\cdot10^{-9}$. Only independent data points were used to calculate this. The number of independent data points is 74, i.e., every third GONG 108-day data set. We use Spearman's $\rho$ for the assessment of the level of correlation of the two quantities as the relation between mode widths and activity is not strictly linear: As can be seen from the right panel in the bottom row in Figure~\ref{fig:3}, the variation of mode widths increases with the level of the $F_{10.7}$ index. However, for values of $F_{10.7}\gtrsim 130$ mode widths appear to be in saturation. The largest discrepancy between the $F_{10.7}$ index and the mode widths can be seen during the time of the camera upgrade in the GONG network around month 60 (in the year 2001). Averaged over all modes, the min-to-max variation of the one-year boxcar smoothed variation of mode widths is 11.5$\pm$0.2\%.
	
	In Figure~\ref{fig:6}, the normalized variation of averages of mode widths (after the corrections for fill and apparent solar radius variation have been applied) for modes of different frequency and harmonic degree ranges are shown as a function of time. Table~\ref{a2:table:1} in Appendix~\ref{app:b} holds information on the number of modes used in each panel of Figure~\ref{fig:6}, the minima and maxima of the parameter variation, the mean uncertainty of each data point, and the correlation of mode widths with the  $F_{10.7}$ index.	
	
	The min-to-max variation of the mode widths over the two observed solar cycles is dependent on mode frequency and harmonic degree. The largest fractional variation is found for modes with $\unit[2400]{\mu Hz}\le \nu \le \unit[3300]{\mu Hz}$ and $31\le l \le 60$ with a min-to-max variation of $26.6\pm0.3\%$. The smallest variation of mode widths is found for the modes in the $\unit[1500]{\mu Hz}\le \nu \le \unit[2400]{\mu Hz}$ and $101\le l \le 150$ parameter regime with a min-to-max variation of $5.5\pm0.2\%$. To suppress the impact of outliers on these min-to-max variations, they are calculated for the one-year boxcar smoothed values of the mode widths. The middle frequency range with $\unit[2400]{\mu Hz}\le \nu \le \unit[3300]{\mu Hz}$ shows the largest min-to-max variation over the solar cycle for each subset of modes of the same range of harmonic degrees. The exception to this is the subset of modes with $101\le l \le 150$, for which the variation over the solar cycle is largest for modes in the high frequency range.
	
	The last two columns of Table~\ref{a2:table:1} give the correlation between the variation of mode widths and the $F_{10.7}$ index as well as the associated two-sided significance value. The highest level of correlation is found for modes with $\unit[1500]{\mu Hz}\le \nu \le \unit[2400]{\mu Hz}$ and $0\le l \le 30$ with a Spearman's rank correlation coefficient of $\rho=0.84$ and a two-sided significance of $p<10^{-10}$. The lowest correlation is found for modes with $\unit[1500]{\mu Hz}\le \nu \le \unit[2400]{\mu Hz}$ and $31\le l \le 60$ with  $\rho=0.31$ and $p\approx0.01$. Within each range of harmonic degrees, the correlation coefficient increases with mode frequency. The exception to this are modes with $0\le l \le 30$, where modes in the middle frequency range show the highest level of correlation with the $F_{10.7}$ index.
	
	\subsection{Mode Amplitudes}\label{sec:3:2}
	In Figure~\ref{fig:4}, the mode amplitudes $A$ are shown as functions of time, fill, and $F_{10.7}$ index (columns) and for three levels of correction (rows) as was previously described for the mode widths (Figure~\ref{fig:3}). The effect of the correction for the change in apparent size of the Sun is more pronounced for mode amplitudes than it is for mode widths. As can be seen in the left panel of the middle row, there is a distinct yearly variation in the mode amplitudes $A_{\text{fill=1, mean}}$ after the fill correction has been applied. This is largely removed in the left panel of the bottom row. The change to the mode amplitudes due to this correction is most obvious through months 110--150, which corresponds to the activity minimum between cycles 23 and 24. In the right panel of the bottom row of Figure~\ref{fig:4}, mode amplitudes $A_{\text{fill=1, radius, mean}}$ are shown as a function of the $F_{10.7}$ index. A clear anti-correlation between the two quantities is visible. For levels of activity up to $F_{10.7}\approx 130$,	the relation appears linear. For higher levels of activity, there is little to no change in mode amplitudes, similar to the behavior seen for mode widths (Figure~\ref{fig:3}). 
	
	\begin{figure}
		\begin{center}
			\includegraphics[width=\textwidth]{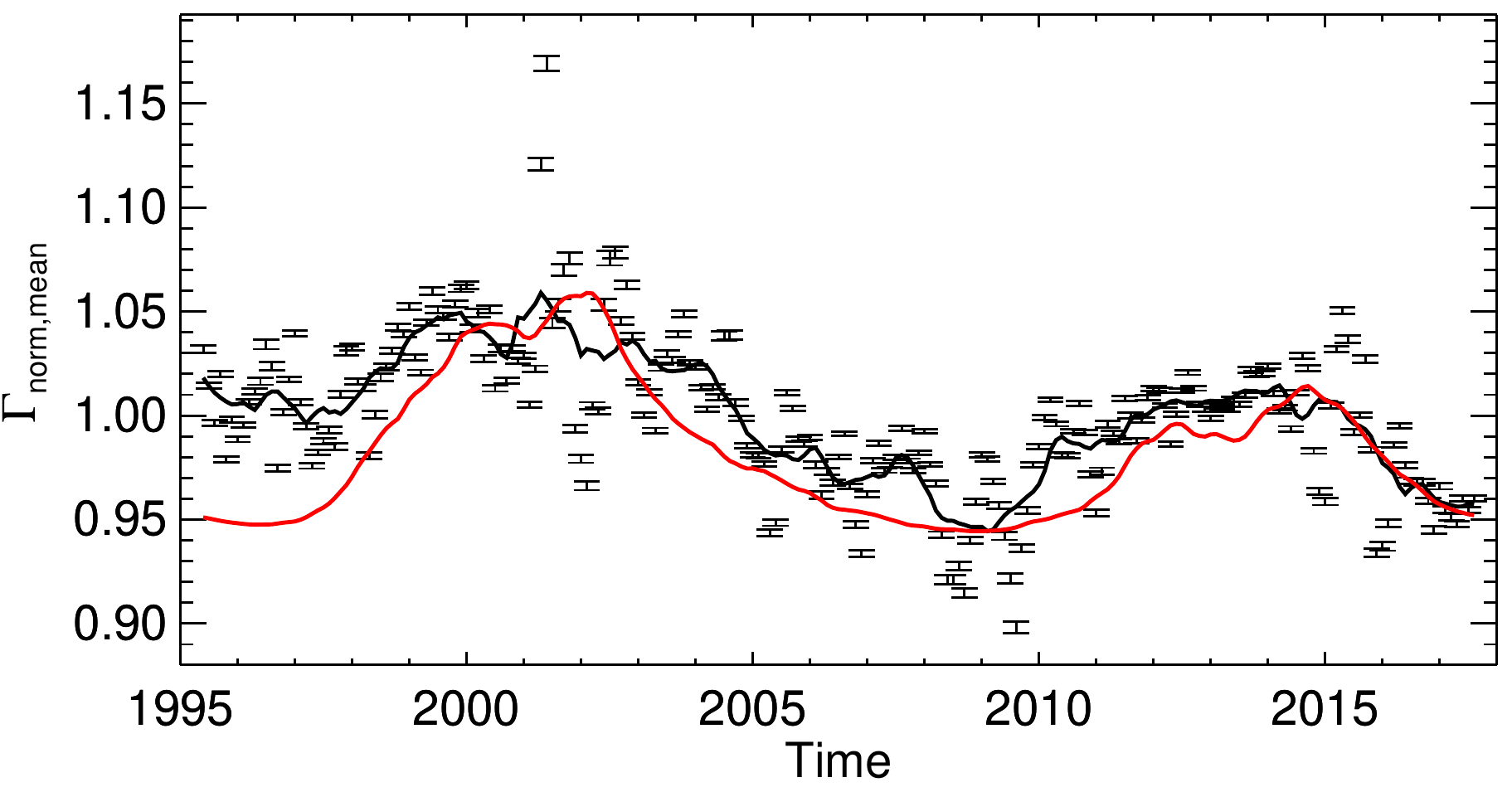}
			\includegraphics[width=\textwidth]{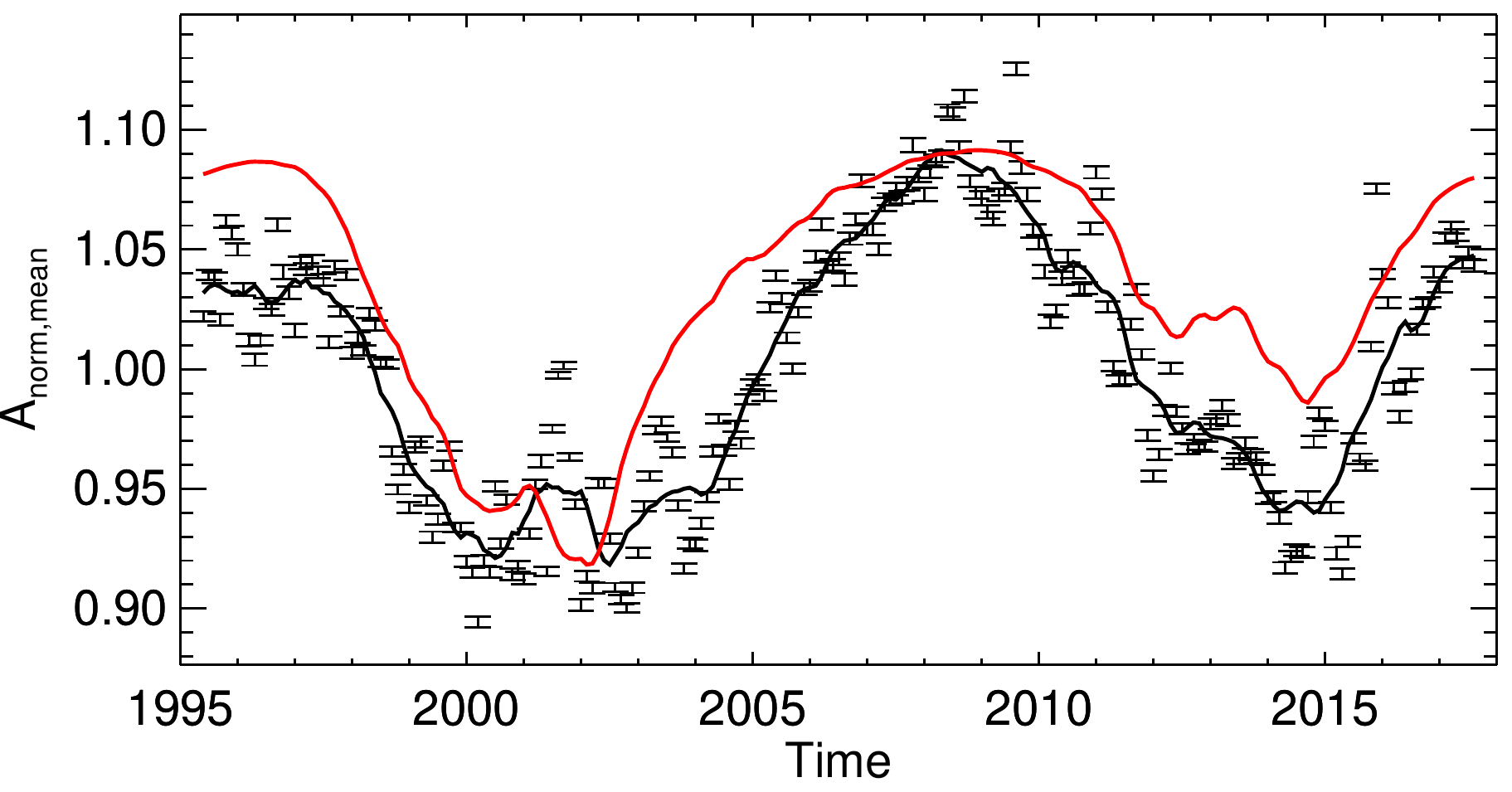}
		\end{center}
		\caption{\textit{Top panel:} normalized mode widths averaged over all modes (black data points). Error bars indicate the error of the mean. The solid black line is the one-year average of the mode widths. The red line is the $F_{10.7}$ solar radio flux as a function of time smoothed by boxcar with a width of one year. It is scaled to match the variation of the mode widths. \textit{Bottom panel:} same as top panel but for mode amplitudes. The $F_{10.7}$ radio flux was flipped for better appreciation of the anti-correlation of the change in mode amplitudes and the level of activity.}
		\label{fig:5}
	\end{figure}
	
	\begin{figure}
		\begin{center}
			\includegraphics[angle=90,width=0.9\textwidth]{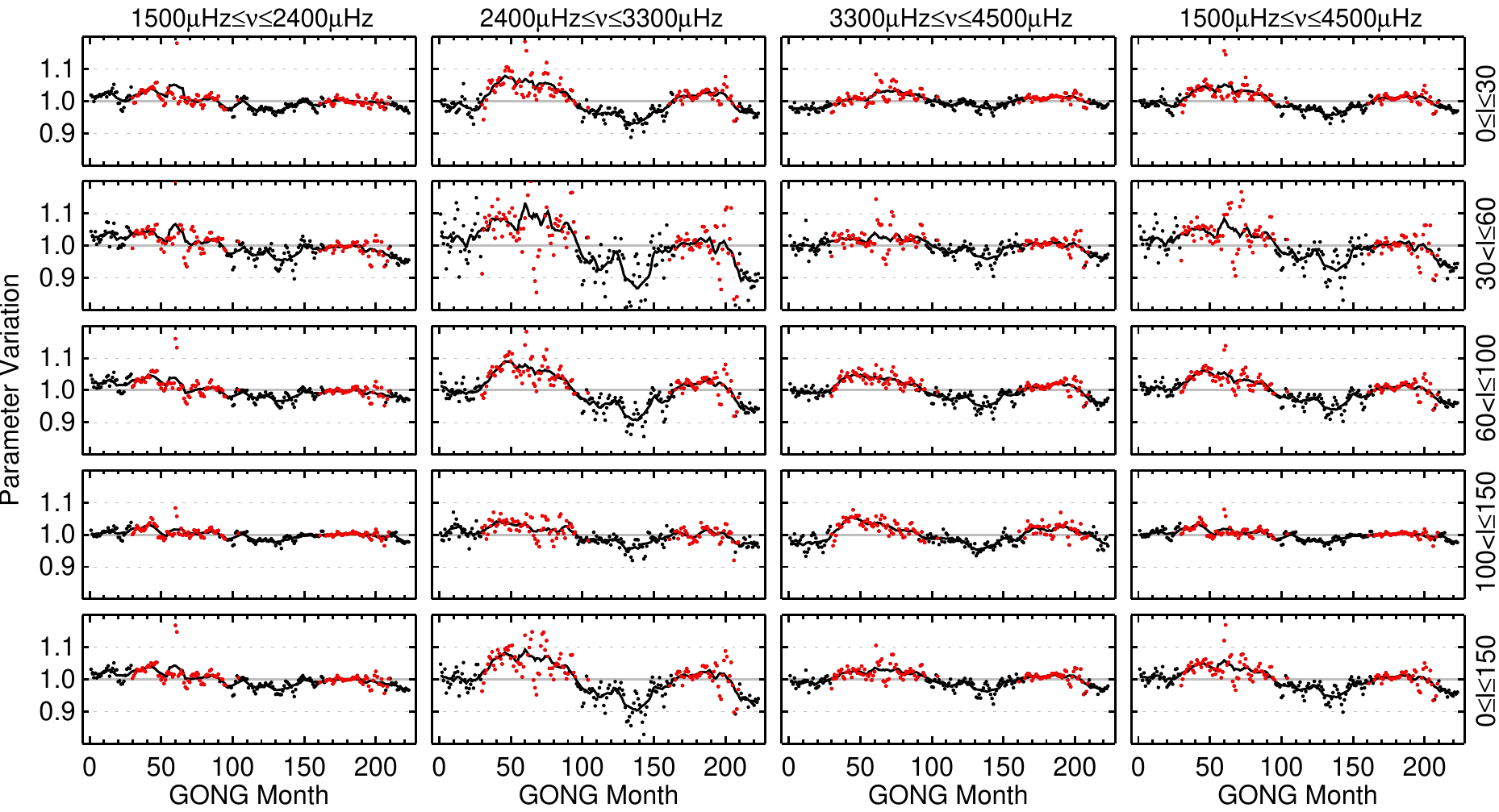}
		\end{center}
		\caption{Temporal variation of mode widths for different ranges of harmonic degrees (rows, harmonic degree range indicated to the right of the fourth column) and mode frequencies (columns, frequency range indicated above the first row). Widths are normalized to the mean for each mode and then averaged over all modes in the respective range of frequency and degree. Months with higher than median $F_{10.7}$ solar radio flux are highlighted by red points. The one-year average is shown by the solid black line. Levels of 1.1 and 0.9 of the mean are indicated by grey dashed lines.}
		\label{fig:6}
	\end{figure}

	\begin{figure}
		\begin{center}
			\includegraphics[angle=90,width=0.9\textwidth]{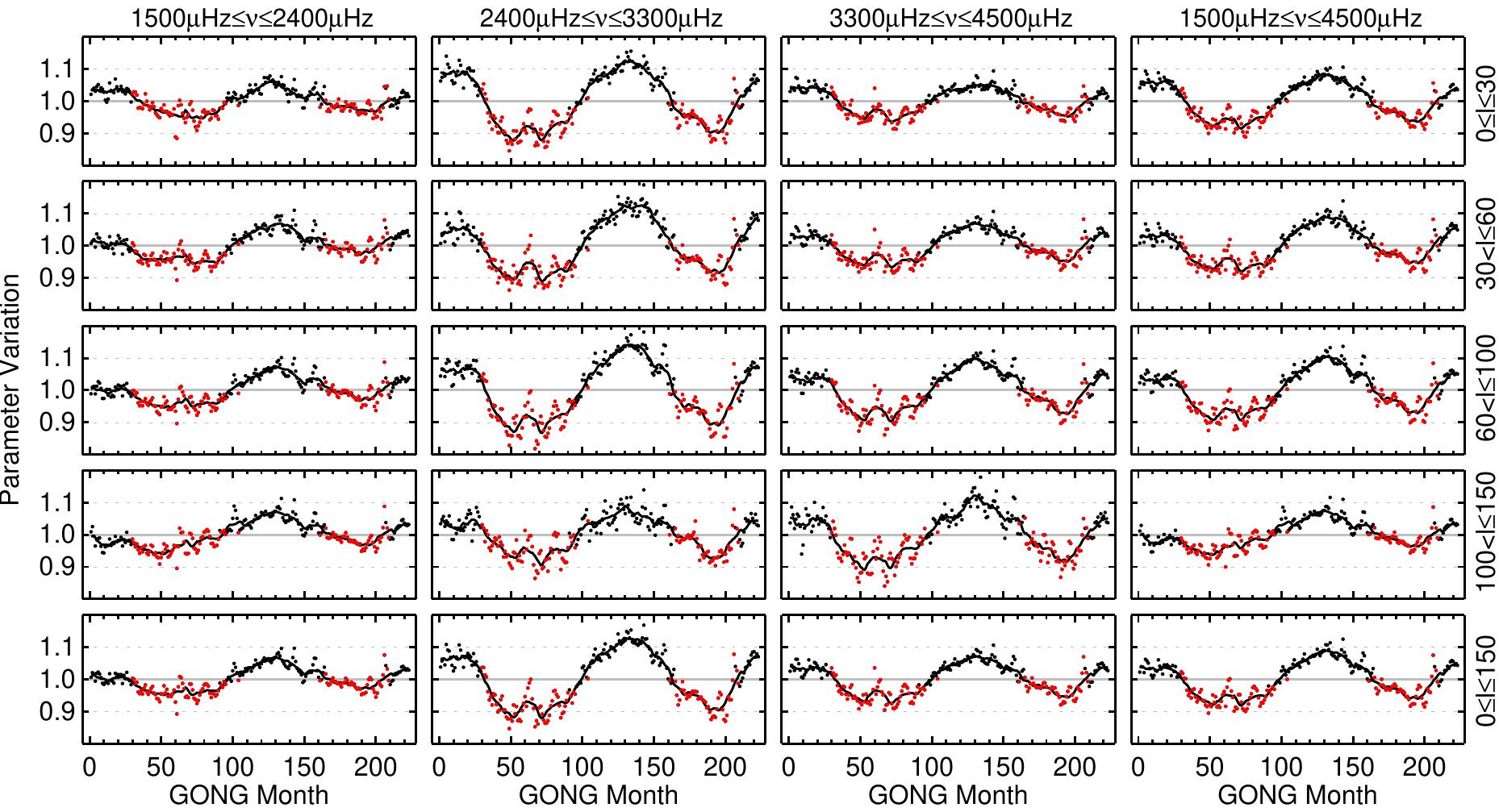}
		\end{center}
		\caption{Same as Figure~\ref{fig:6} but for mode amplitudes.}
		\label{fig:7}
	\end{figure} 
	
	The normalized, averaged, and corrected temporal variation for mode amplitudes is shown in the lower panel of Figure~\ref{fig:5}. The last row of Table~\ref{a2:table:2} in Appendix~\ref{app:b} gives more detailed information about the number of modes within each parameter range and the error bars presented in this plot. The error bars in Figure~\ref{fig:5} are computed as the standard error of the mean. The solid lines are the data (black) and $F_{10.7}$ index (red) smoothed with a one-year boxcar. The $F_{10.7}$ index is scaled to match the min-to-max variation of the mode amplitudes and flipped for better appreciation of the anti-correlation of the two quantities. Averaged over all modes, the min-to-max variation is 17.3$\pm$0.2\%. The correlation coefficient (Spearman's rank correlation coefficient) between amplitudes and the $F_{10.7}$ index (both unsmoothed) is $\rho=-0.91$ with $p<10^{-10}$. Only independent data points were used in the computation of the correlation coefficient.
	
	The equivalent of Figure~\ref{fig:6}, where the normalized variation of mode widths is shown for different mode parameter regimes, is presented in Figure~\ref{fig:7} for mode amplitudes. Detailed information on the individual panels is given in Table~\ref{a2:table:2} in Appendix~\ref{app:b}. The largest variation is found for modes with $\unit[2400]{\mu Hz}\le \nu \le \unit[3300]{\mu Hz}$ and $61\le l \le 100$ with a min-to-max variation of $27.4\pm 0.4\%$. Modes with frequencies $\unit[1500]{\mu Hz}\le \nu \le \unit[2400]{\mu Hz}$ and harmonic degrees $0\le l \le 30$ exhibit the smallest variation over the 22 years of data with a fractional change of $11.6\pm0.5\%$. The anti-correlation between the level of activity and the change in mode amplitudes is highest for modes with $\unit[2400]{\mu Hz}\le \nu \le \unit[3300]{\mu Hz}$ and $0\le l \le 30$ with a rank correlation of $\rho=-0.94$ and $p<10^{-10}$. For the three ranges of harmonic degrees ($l$ between 0--30, 31--60, and 61--100) the largest variation is found in the middle frequency range (2400--3300$\unit{\mu Hz}$). For modes with $101\le l \le 150$, this is observed for the high frequency modes.
	
\begin{figure}[t]
		\begin{center}
			\includegraphics[width=\textwidth]{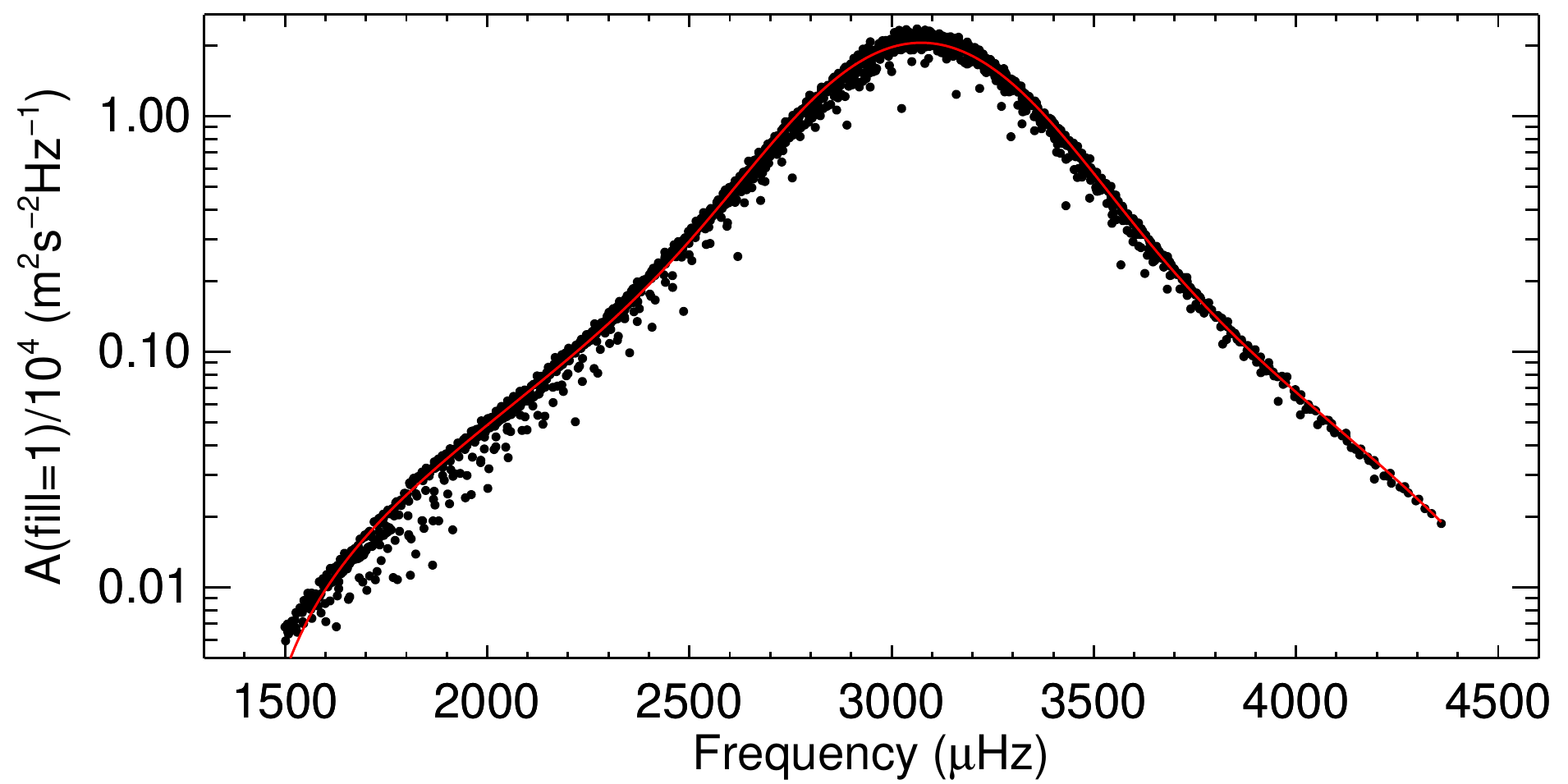}
			\includegraphics[width=\textwidth]{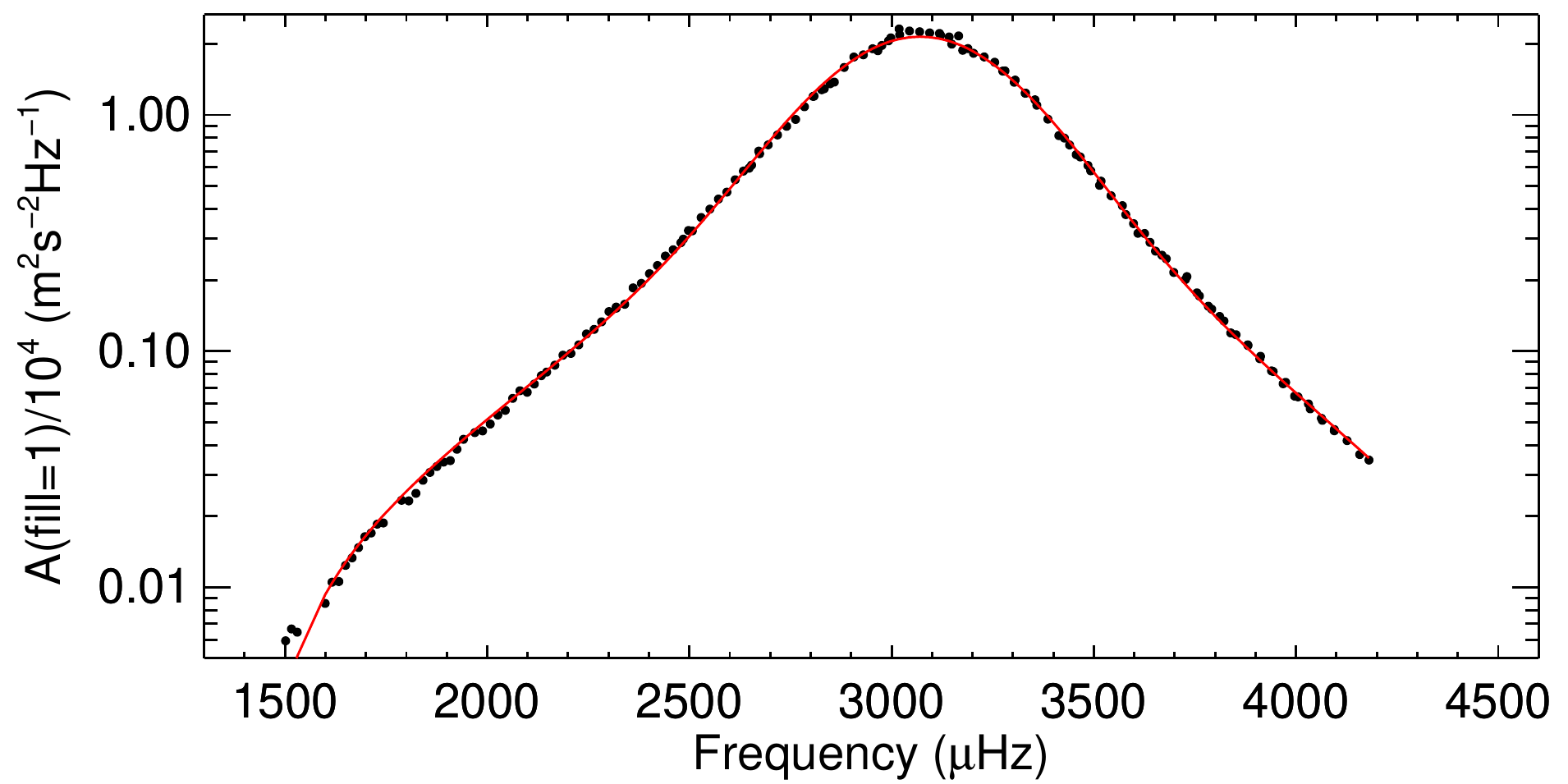}
		\end{center}
		\caption{Mode amplitudes as a function of mode frequency. Amplitudes are averaged over all GONG months after correction for fill factor. The red curve is an asymmetric Voigt profile fitted to the data. \textit{Top panel}: amplitudes of modes with $2\le l \le 150$. \textit{Bottom panel:} amplitudes of modes with $31\le l \le 40$}. 
		\label{fig:8}
\end{figure}
	
	\subsection{Frequency Distribution of Mode Amplitudes}\label{sec:3:3}
	In the top panel of Figure~\ref{fig:8}, the amplitudes of all modes with $2\le l \le 150$, which are present in at least 50\% of the GONG months, are shown as a function of mode frequency on a logarithmic ordinate divided by $10^4$. Here, the amplitudes interpolated to $\text{fill}=1$ are used. Any variation due to the residual change in apparent solar radius is averaged out as the data cover 22 years. By lowering the presence rate to 50\%, more modes, especially of $l\gtrsim 100$, are included and the mode amplitudes can be investigated for separate ranges of harmonic degrees (see the bottom panel of Figure~\ref{fig:8} for an example). This increases the total number of included modes from 1272 to 1543.
		
	The solid red lines in Figure~\ref{fig:8} are fits of asymmetric Voigt profiles to the mode amplitudes. This profile is described by
	\begin{align}\label{eq:1}
	V(\nu) = \mathcal{A}\cdot \left(b+ a\cdot\int\limits_{-\infty}^{\infty}G(x) L(\nu-x) \D x\right)
	\end{align}
	where $b$ is an offset, $a$ is a factor to adjust the height of the profile, and the integral extends over the entire frequency axis with frequencies $x$. The Gaussian function $G(\nu)$ and the Lorentz function $L(\nu)$ in Equation~(\ref{eq:1}) are given by
	\begin{align}
	G(\nu)&=\frac{\exp(-\nu/2\sigma^2)}{\sigma\sqrt{2\pi}},\label{eq:2}\\
	L(\nu)&=\frac{\gamma}{\pi\left(\nu^2+\gamma^2\right)},\label{eq:3}
	\end{align}
	with the standard deviation of the Gaussian $\sigma$ and the half-width at half maximum of the Lorentzian $\gamma$. The asymmetry is introduced with the function
	\begin{align}\label{eq:4}
	\mathcal{A} = \frac{1}{\pi} \left(\tan^{-1}\left(S\cdot\left(\nu - \nu_{\text{max}}\right)/\Sigma\right) + 0.5\right),
	\end{align}
	where $S$ is the asymmetry parameter, $\nu_{\text{max}}$ is the frequency of the maximum of the symmetric Voigt profile, and the full-width-at-half-maximum of the Voigt profile, $\Sigma$, can be approximated by \citep{Whiting1968,Olivero1977}
	\begin{align}\label{eq:5}
	\Sigma = 1.0692 \gamma+ \sqrt{0.86639 \gamma^2 + 8\ln(2)\sigma^2}.
	\end{align}
	
	The resulting fit parameters for amplitudes of modes with $2\le l \le 150$ are given in Table~\ref{table:1}. For the frequency of maximum amplitude, we find $\nu_{\text{max}}=\unit[3079.76\pm0.17]{\mu Hz}$. The width of the Voigt profile is $\Sigma=\unit[611.8\pm0.5]{\mu Hz}$. The parameter $S=-0.100\pm0.002$ indicates that the distribution is slightly left-tailed (left-skewed, right-leaning). The $\chi^2_{\text{red}}$ of the fit is $32.8$. This rather high $\chi^2_{\text{red}}$ value is due to the spread in the distribution of the mode amplitudes for modes of different harmonic degree and the small error bars. Other profiles were tested (pure Gaussian, pure Lorentzian, both Gaussian and Lorentzian with asymmetry included), but the asymmetric Voigt yielded the best $\chi^2_{\text{red}}$. We excluded modes with $l=0,1$ because their amplitudes are less well defined and including them here would increase the $\chi^2_{\text{red}}$ to $\approx135$.
	
	It should be noted that the maximum of the mode amplitudes $\nu_{\text{max}}$ presented here is not exactly what is often referred to as $\nu_{\text{max}}$ in the literature. It is usually measured from Sun-as-a-star data which only include low harmonic degrees up to $l\lesssim 4$ \citep{Broomhall2009a}. Here, $\nu_{\text{max}}$ is the maximum of the mode amplitudes in the frequency spectrum of the spherical harmonic transform of the GONG Doppler velocity data for all harmonic degrees up to $l=150$. Typical values for $\nu_{\text{max}}$ include:\label{c6:page:numax} $\unit[3050]{\mu Hz}$ by \citet{Kjeldsen1995}, $\unit[3104]{\mu Hz}$ as a calibrated value to obtain better results from asteroseismic scaling relations by \citet{Mosser2013}, and $\unit[3120]{\mu Hz}$ from SOHO/Virgo data by \citet{Kallinger2010}. If $\nu_{\text{max}}$ is measured from the velocity of the modes (as will be discussed in Section~\ref{sec:4}, the mean velocity of the modes is proportional to mode width $\Gamma$ times mode amplitude $A$), the damping widths $\Gamma$ are included in the measured quantity. As mode widths change with mode frequency and harmonic degree (see next section), the maxima in mode amplitude $A$ and mean velocity of the modes (proportional to $\Gamma\cdot A$) occur for slightly different frequencies.
	
	We also investigated the mode amplitudes for smaller ranges of harmonic degrees. For this, we separated the modes into groups of typically ten harmonic degrees. As an example, we show the amplitudes of modes with $31\le l \le 40$ (black data points) and the fit of an asymmetric Voigt profile (solid red line) in the lower panel of Figure~\ref{fig:8}. As can be seen from Figure~\ref{fig:8}, the frequency dependence of the amplitudes of modes of similar harmonic degree is well captured by the asymmetric Voigt profile. The resulting fit parameters for this and more ranges of harmonic degrees are presented in Table~\ref{a3:table:1} in Appendix~\ref{app:c}. There, we also show figures of the change of the fit parameters with harmonic degree. Apart from the width of the Voigt profile $\Sigma$, all parameters show systematic variations with mode degree. The frequency of maximum amplitude $\nu_{\text{max}}$ has a maximum value for intermediate degree modes and decreases substantially at low and high degrees (below $l=30$ and above $l=90$). The skewness $S$ is anti-correlated with $\nu_{\text{max}}$. The width $\Sigma$ is approximately constant except for the band of highest harmonic degrees considered. Amplitudes first increase with harmonic degree, reach a maximum around $l\approx40$, and then decreases again.

\begin{table}
	\caption{Parameters of fit to the frequency distribution of mode amplitudes for modes with $2\le l \le 150$.}\label{table:1}
	\begin{tabular}{cccc}
		\hline 
		$\nu_{\text{max}}$ & $\sigma$&$\gamma$ & $\Sigma$ \\ 
		\text{[}$\mu$Hz]& [$\mu$Hz]&[$\mu$Hz]&[$\mu$Hz]\\
		\hline
		3079.76$\,\pm\,$0.17 & 181.8$\,\pm\,$0.3 &150.9$\,\pm\,$0.2 & 611.8$\,\pm\,$0.5 \\ 
		\hline 
		&&&\\
		\hline
		$a/10^4$& $b$ & $S$ &$\chi^2_{\text{red}}$\\ 
		\text{[}$\unit{m^2\,s^{-2}\,Hz^{-1}}$] & [$\unit{m^2\,s^{-2}\,Hz^{-1}}$]  & &\\
		\hline
		3299$\,\pm\,$2& -581$\,\pm\,$1 & -0.100$\,\pm\,$0.002  &32.8\\ 
		\hline 
	\end{tabular} 
\end{table}
	
\section{Results for Physical Quantities}\label{sec:4}
	The amplitude $A$ is given as power per frequency bin $\unit{m^2\,s^{-2}\,Hz^{-1}}$. Hence, calculating the product of mode width and mode amplitude $\Gamma_{nl}\cdot A_{nl}$ has the unit of squared velocity. The mean squared velocity of the modes can be calculated as \citep{Goldreich1994}
	\begin{align}
	\langle v^2_{nl}\rangle = \frac{\pi}{2} C_{\text{vis}} \Gamma_{nl} A_{nl}. \label{eq:6}
	\end{align}
	Symmetric Lorentzian profiles are fitted to the peaks in the power spectrum by the GONG pipeline. This is taken into account by the scaling factor $\pi/2$. The quantity $\Gamma_{nl}\cdot A_{nl}$ is referred to as the \textit{mode area}, i.e., the area under the fitted peak in the spectrum \citep{Komm2000}. The factor $C_{\text{vis}}=3.33$ corrects for the reduced visibility of modes due to leakage effects \citep{Hill1998}. 
	
	Due to the different cavities in which modes of different harmonic degree and frequency propagate, the fraction of the mass of the Sun that is affected by different modes varies \citep[see, e.g.,][]{Basu2016}. The \textit{mode mass} $M_{nl}$ is given by
	\begin{align}
	M_{nl} = M_{\odot} I_{nl}, \label{eq:7}
	\end{align}
	where $M_{\odot}$ is the solar mass and the mode inertia is calculated as \citep{Christensen-Dalsgaard1991b}
	\begin{align}
	I_{nl} = \frac{4\pi \int_0^{R_{\odot}} \rho(r)\left({\xi^r_{nl}(r)}^2 + l(l+1){\xi^h_{nl}(r)}^2\right)r^2\D r}{M_{\odot} \left({\xi^r_{nl}(R_{\odot})}^2 + l(l+1){\xi^h_{nl}(R_{\odot})}^2\right)} \,. \label{eq:8}
	\end{align}
	Here, $\xi^r_{nl}$ and $\xi^h_{nl}$ are the radial and horizontal displacement eigenfunctions associated with the oscillation, $\xi_{nl}^r(R_{\odot})$ and $\xi_{nl}^h(R_{\odot})$ are their values at the photospheric radius, and $\rho$ is mass density. We calculate mode inertias from the eigenmodes of the standard solar model `Model S' \citep{Christensen-Dalsgaard1996}.
	
	The energy that is stored in the modes can be calculated as the product of mode mass (Equation~\ref{eq:7}) and mean squared velocity (Equation~\ref{eq:6}):
	\begin{align}
	E_{nl} = M_{nl}	\langle v^2_{nl}\rangle. \label{eq:10}
	\end{align} 
	This is the total mode energy, i.e., the sum of kinetic and potential energy \citep{Goldreich1994}. The rate at which energy is supplied to the modes can be calculated by the product of the energy in the modes and the radian damping rate of the modes $2\pi\Gamma_{nl}$ \citep{Kumar1988,Goldreich1994}:
	\begin{align}
	\frac{\D E_{nl}}{\D t} = 2\pi E_{nl} \Gamma_{nl} = \pi^2 C_{\text{vis}} M_{nl} A_{nl} \Gamma_{nl}^2,\label{eq:11}
	\end{align}
	which is proportional to the product of squared mode width $\Gamma_{nl}^2$ and mode amplitude $A_{nl}$.
	
	\begin{figure}
		\begin{center}
			\includegraphics[width=\textwidth]{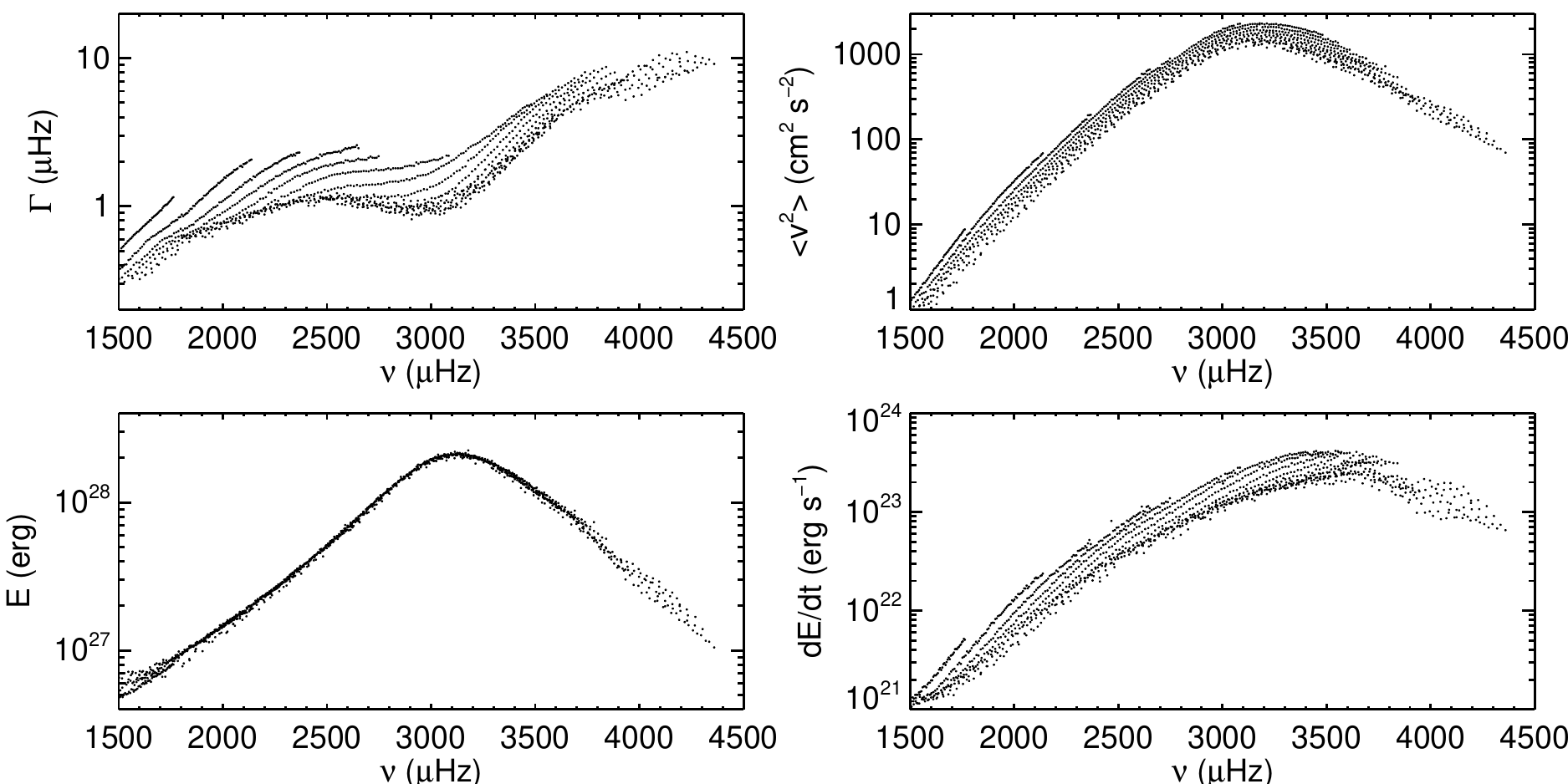}
		\end{center}
		\caption{Mode widths $\Gamma$, mean squared velocity $\langle v^2\rangle$, mode energies $E$, and energy supply rates $\D E/\D t$ as functions of frequency. Averaged over all GONG months. }
		\label{fig:10}
	\end{figure}
	
	In Figure~\ref{fig:10}, the mode width (damping rate) $\Gamma$, mean squared velocity $\langle v^2_{nl}\rangle$, mode energy $E$, and energy supply rate $\D E/\D t$ of modes with harmonic degree $l>10$ are shown on logarithmic scales as functions of mode frequency. Modes with $l\le10$ are excluded here to further reduce the scatter in these plots. The parameters of these modes are less well-defined due to their small number of azimuthal components, see Section~\ref{sec:2:2}. For each mode, the values for $\Gamma$, $A$, $\Gamma \cdot A$, and $\Gamma^2 \cdot A$ are calculated as the mean over all time samples after correction for the fill factor. Any effects from the activity cycle or residual yearly variations are averaged out as the data set spans two complete Schwabe cycles. The conversion of these quantities to physical units is then done according to Equations~(\ref{eq:6}),  (\ref{eq:10}), and (\ref{eq:11}).
	
	The mode widths (upper left panel in Figure~\ref{fig:10}) increase from 0.3--0.5\,$\unit{\mu Hz}$ at low mode frequencies $\approx\unit[1500]{\mu Hz}$ to values between 1--2\,$\unit[]{\mu Hz}$ for mode frequencies between 2400--3000\,$\unit{\mu Hz}$. For higher mode frequencies, mode widths increase again, reaching $\Gamma\approx\unit[10]{\mu Hz}$ for the highest mode frequencies. The mean squared velocity (upper right panel) increases monotonically with mode frequency until it reaches its maximum for modes with frequencies of $\approx\unit[3200]{\mu Hz}$. The maximum mean velocity value is $v\approx\unit[37]{cm\,s^{-1}}$. The mode widths, mean squared velocity, as well as the energy supply rate (lower right panel) show ridges of modes with equal radial order. Different ridges are slightly offset from one another but show the same behavior with frequency. The ridge structure is not apparent in the mode energies (lower left panel). The difference between mode energy $E$ and the other three quantities is the inclusion of mode inertia, see Equations~(\ref{eq:6}), (\ref{eq:7}), (\ref{eq:10}), and (\ref{eq:11}). 
	
\begin{figure}
		\begin{center}
			\includegraphics[width=\textwidth]{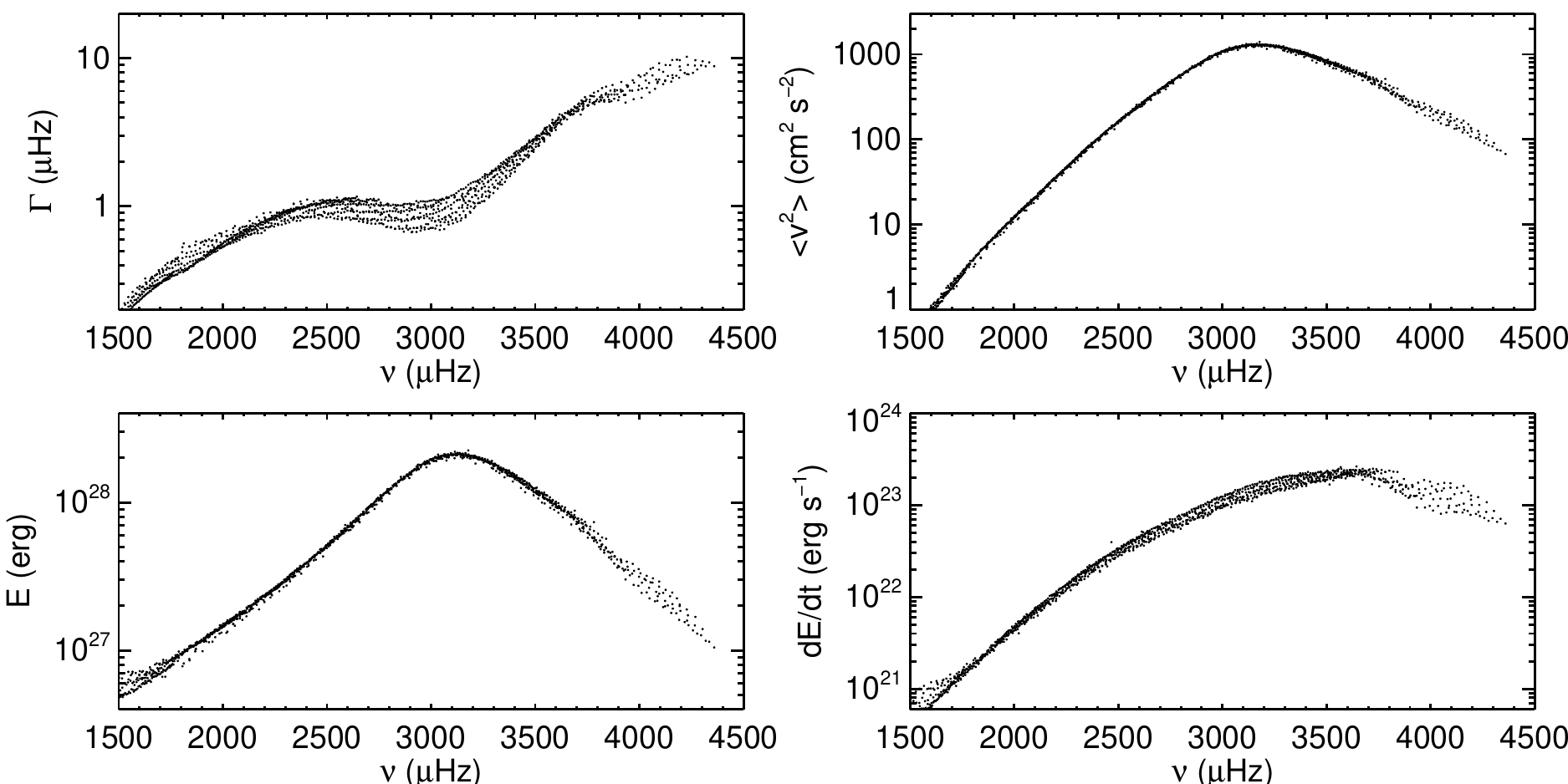}
		\end{center}
		\caption{Same as Figure~\ref{fig:10} but normalized for mode inertia with Equation~(\ref{eq:12}). Mode energy $E$ is unchanged.}
		\label{fig:11}
	\end{figure}

	A normalized mode inertia can be calculated by computing the ratio of the mode inertia $I_{nl}$ to the inertia of radial modes $\overline{I_{n0}}$ interpolated to the frequency of the mode $\nu_{nl}$ \citep{Christensen-Dalsgaard1996, Aerts2010}:
	\begin{align}
	Q_{nl} = \frac{I_{nl}}{\overline{I_{n0}}\left(\nu_{nl}\right)}.\label{eq:12}
	\end{align}
	The result of multiplying mode widths, mean squared velocities, and energy supply rates with  $Q_{nl}$ is shown in Figure~\ref{fig:11}.
	The overall frequency dependence of the quantities is unchanged. However, ridges of different radial orders are now largely collapsed into one. This can best be observed for $\langle v^2_{nl}\rangle$ (top right panel), which is only composed of one thin line of data points without any residual ridge structure. The steep gradient in mode inertia at low frequencies is not perfectly represented by scaling $Q_{nl}$ with $\overline{I_{n0}}$. Some of the residual scatter in mode widths, energies, and energy supply rate at low mode frequencies is due to this. Different scalings, e.g., with the interpolated inertia of $l=50$ modes, give somewhat different results \citep{Komm2000}. However, the scatter in the scaled quantities is never completely removed. The remaining dependence observed in mode widths across all frequencies may hold information on a degree dependence of the involved damping mechanisms.
	
	To be able to appreciate the temporal change of the four quantities (mode width, mean squared velocity, mode energy, mode energy supply rate), the parameters of many modes have to be averaged. As the quantities cover two to three orders of magnitude for the investigated set of modes, the frequency range of averaging has to be restricted. Otherwise, values which differ by orders of magnitude would contribute to the average. In Table~\ref{table:2}, the frequency ranges of the modes which are taken for the averaging of the four quantities, and the number of modes within these frequency ranges are given. The frequency ranges are chosen to select as many modes as possible with similar parameter values (within about a factor of two). 
	
\begin{figure}
		\begin{center}
			\includegraphics[width=\textwidth]{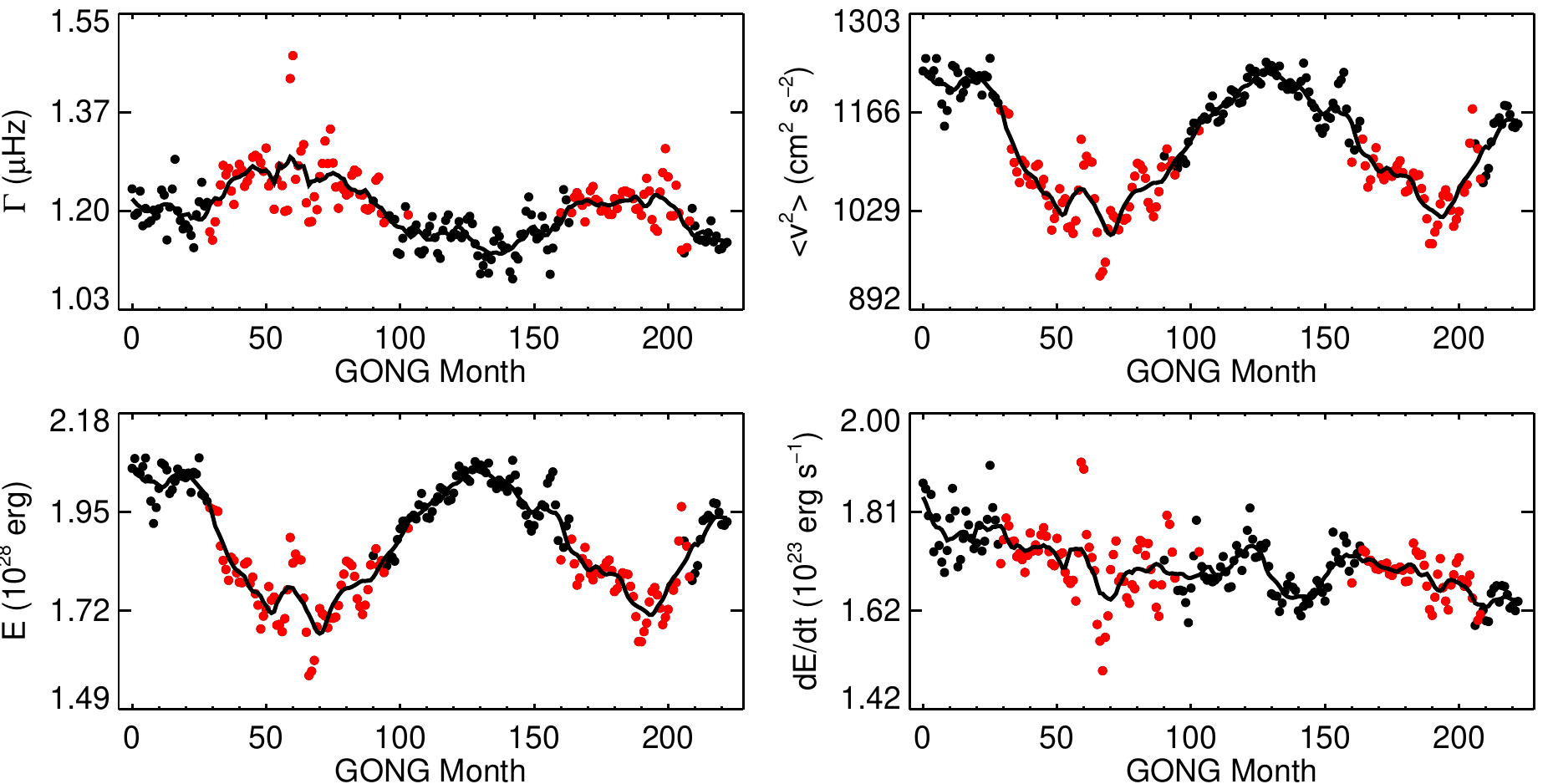}
		\end{center}
		\caption{Mode widths $\Gamma$, mean squared velocity $\langle v^2\rangle$, mode energies $E$, and energy supply rates $\D E/\D t$ as functions of time. Averaged over all modes in the frequency ranges given in Table~\ref{table:2}. }
		\label{fig:12}
	\end{figure}

	The result of this averaging is presented in Figure~\ref{fig:12} as a function of time. The red data points indicate times of higher than median level of activity and the solid black lines are the data smoothed with a one-year boxcar. The last two rows of Table~\ref{table:2} give the rank correlation of independent data points between the four quantities and the $F_{10.7}$ index as well as the two-sided significance value. As before, the mode widths are correlated with the level of magnetic activity with $\rho=0.69$. Both, the temporal variation in mean squared velocity and mode energy are highly anti-correlated with the level of solar activity with a value of $\rho=-0.88$. The energy supply rate is not correlated with solar activity for the set of modes investigated here as $\rho=-0.01$ and $p=0.88$.
	
	\section{Summary and Discussion}\label{sec:5}
	This study is the first to investigate the activity-related p-mode parameter changes of resolved data covering an complete solar magnetic cycle. We analyzed mode parameter data from the ground-based GONG network. Systematic effects due to the spatial masking of the GONG full-disk solar images, the imperfect temporal sampling, and a yearly modulation due to changes in the apparent radius of the Sun were corrected for. Measurements of mode widths and mode amplitudes are generally less robust than measurements of mode frequencies. Thus, averages over azimuthal orders as well as modes of different ranges of frequencies and harmonic degrees have to be performed in order to reach meaningful results for the temporal variation of these mode parameters. 
	
	Little attention has been given to mode widths and amplitudes (for resolved solar data; \citealp{Komm2000, Korzennik2013, Korzennik2017} are the most notable exceptions), let alone their temporal variation over the solar cycle. The reason for this is that even without incorporating solar cycle variations damping widths and amplitudes cannot be accurately modelled with the existing theory of convection and mode excitation (see, e.g., \citealp{Houdek1999,Houdek2015,Basu2016}). Thanks to the long, uninterrupted GONG time series that is now available, we were now able to investigate the variation of the mode widths and amplitudes as functions of the level of solar activity for different subsets of mode frequencies and harmonic degrees. When the last studies similar to the present one were published, about three years' \citep{Komm2000a} and four years' \citep{Komm2000} worth of GONG data had been recorded.
	
	\begin{table}
		\caption{Frequency ranges for averaging of physical quantities, number of modes in this frequency range, and correlation coefficients with the $F_{10.7}$ index.}\label{table:2}
		\begin{tabular}{ccccc}
			\hline 
			& $\Gamma$& $\langle v^2\rangle$ & $E$ & $\D E/\D t$  \\ 
			\hline 
			Frequency range [$\unit{\mu Hz}$]& $\unit[2400]{}$--$\unit[3000]{}$& $\unit[2900]{}$--$\unit[3300]{}$  & $\unit[2900]{}$--$\unit[3300]{}$  & $\unit[3000]{}$--$\unit[3600]{}$ \\ 
			Number of modes&	358 & 237 & 237 & 385\\
			Correlation $\rho$ &0.69&-0.88&-0.88&-0.01\\
			p value &$<10^{-10}$&$<10^{-10}$&$<10^{-10}$&0.88\\
			\hline 
		\end{tabular} 
	\end{table}
	
	For the mode widths, which were averaged over 1275 modes, we found a variation of 11.5$\pm0.2\%$ between the minimum and maximum level of activity in the investigated time period. Mode amplitudes varied by 17.3$\pm 0.2\%$ over the same time (averaged over 1272 modes). Overall, the results from previous analyses of the variation of mode widths and amplitudes with the level of solar activity by, e.g., \citet{Komm2000a}, \citet{Komm2000}, and \citet{Broomhall2015} could be confirmed. However, we find a larger variation of mode widths and amplitudes than was expected by \cite{Komm2000a}. By extrapolating the fractional change per Gauss of widths and amplitudes to the full minimum-to-maximum change of activity over a solar cycle, they expected widths to vary by 7\% and amplitudes to vary by 16\% averaged over all modes. Thus, the changes of mode widths we report here are about two thirds larger than expected by \cite{Komm2000a}, while the changes of mode amplitudes agree within a few percent. The smaller estimated variation of mode widths by \cite{Komm2000a} was most likely due to the shorter time series and differences in which modes they included.
	
	Like \cite{Komm2000a}, we too investigated the change of mode widths and amplitudes for subsets of mode frequencies and harmonic degrees. While they averaged over $\unit[100]{\mu Hz}$ or five harmonic degrees (their Figures 10 and 11), we did this for larger ranges in frequency and harmonic degree, see Figures~\ref{fig:6}--\ref{fig:7} and Tables~\ref{a2:table:1}--\ref{a2:table:2}. We can confirm their findings that the change of mode widths and amplitudes with activity is largest for modes around the frequency of maximum amplitude. We also can confirm that the change is more strongly dependent on mode frequency than it is on harmonic degree: for example, the amplitude of the fractional variation of mode amplitudes for modes with $61\le l \le 100$ is 12.9\%, 27.4\%, and 19.6\% for the low-, mid-, and high-frequency ranges, respectively. Keeping the frequency range the same, the fractional variation of mode amplitudes for modes with $2400\le \nu\le 3300$ is 24.8\%, 23.7\%, 27.4\%, and  18.4\% for the four ranges of harmonic degrees in ascending order.

	The amplitude of the shifts of mode frequencies with the level of solar activity is known to increase with mode frequency \citep{Jimenez-Reyes1998, Salabert2015a}. In contrast, this behaviour is observed neither for mode widths nor for mode amplitudes. For these two parameters, the largest variation over the solar cycle is observed around the frequency of maximum amplitude $\nu_{\text{max}}$, see Tables~\ref{a2:table:1} and \ref{a2:table:2}, which was already measured by \citealp{Komm2000}, see their Table 3. This frequency dependent variation is in agreement with the theoretical calculations by \cite{Houdek2001}, who found that mode damping widths for frequencies between 2500--3000\unit{$\mu$Hz} increase with decreasing characteristic size of granulation, i.e., with higher levels of magnetic activity. 
	
	Thus, the mechanisms through which the frequencies are changed by the presence of the magnetic field associated with the solar cycle differ from those physical mechanisms which perturb the mode damping widths and amplitudes. Mode widths and amplitudes are determined by the excitation and damping of the acoustic oscillations in very shallow layers, where convection is most vigorous \citep{Balmforth1992, Rimmele1995, Houdek2015}. Hence, this is where the changing magnetic field over the solar cycle affects mode widths and amplitudes most strongly. In contrast to this, mode frequencies can be perturbed by magnetic fields which are located much deeper in the Sun (see, e.g., \citealp{Gough1990}). An analysis of the timing of the changes in mode frequencies, amplitudes, and widths may yield information on the evolution and possibly the gradual ascent of magnetic field concentrations through the outer part of the convection zone. 
	
	The mode amplitudes as a function of mode frequency were fitted with an asymmetric Voigt profile. From this, we found a frequency of maximum amplitude at $\nu_{\text{max}}=\unit[3079.76\pm0.17]{\mu Hz}$. The frequency of maximum amplitude $\nu_{\text{max}}$ is of great interest to asteroseismic studies due to its use in scaling relations for stellar mass and radius. The value found here compares well to those found in the literature (see the remarks on page \pageref{c6:page:numax}). Mode amplitudes are a function of radial order and frequency, as can be seen in the top panel of Figure~\ref{fig:8}. The spread in the distribution of the mode amplitudes as a function of mode frequency could not be removed by multiplication with mode inertia as for, e.g., mean squared velocity of the modes. This indicates that the excitation of modes is intrinsically not only a function of mode frequency. We find that amplitudes of modes of similar harmonic degree follow the asymmetric Voigt profile very well (bottom panel of Figure~\ref{fig:8}) and that the fit parameters vary with harmonic degrees of the modes, see discussion in Appendix~\ref{app:c}.
	
	The physical quantities of mean squared velocity, mode energy, and energy supply rate were calculated from the mode widths and amplitudes. Ridges of the same radial order are evident in the mode widths, mean squared velocities, and the energy supply rates (Figure~\ref{fig:10}). This is removed by scaling them with their mode inertia $Q_{nl}$ (Figure~\ref{fig:11}), indicating that $Q_{nl}$ is the correct scaling for mode widths, mean squared velocities, and energy supply rates \citep{Komm2000}. The maximum of the velocity of the modes is approximately $\unit[37]{cm\,s^{-1}}$. This value is somewhat higher than the value of $\approx\unit[27]{cm\,s^{-1}}$ found by \citet{Chaplin1998a} for radial modes from BiSON data. As modes of very low harmonic degree (below $l=10)$ are not included in this investigation, it may well be that the velocity of these modes would be closer to the value of \citet{Chaplin1998a} than that of modes for higher degree. This is supported by the fact that, on closer inspection, modes of lower harmonic degree are concentrated at the lower edge of the distribution shown in the top right panel of Figure~\ref{fig:10}.
	
	To obtain the temporal variation of these quantities (i.e., mean squared velocity, energy, and energy supply rate; in physical units, not normalized to their temporal mean), an average over modes within a frequency range specific to each quantity (see Table~\ref{table:2}) was calculated for the inertia corrected parameter values. 
	
	The average of the mode widths was found to vary between a maximum width of $\unit[1.36\pm0.01]{\mu Hz}$ at the maximum of solar cycle 23 and a minimum of $\unit[1.19\pm0.01]{\mu Hz}$ during the activity minimum between cycles 23 and 24. The mean squared velocity varied between extrema of $\unit[1065\pm2]{cm^2\, s^{-2}}$ and $\unit[1249\pm4]{cm^2\,s^{-2}}$. The mean mode energy exhibits variation between $\unit[1.64\pm0.01\times10^{28}]{erg}$ and $\unit[2.06\pm0.01\times10^{28}]{erg}$. This highlights that detectability of solar-like oscillation is reduced in stars with high levels of magnetic activity, as has been investigated and shown by \citet{Chaplin2011}. These fractional changes are considerably different for modes of different frequencies, as can be seen from Figures~\ref{fig:6} and \ref{fig:7}. Qualitatively, our results agree with those of, e.g., \cite{Komm2000}, \cite{JimenezReyes2003}, and \cite{Salabert2006}. Due to the differences in the included mode degrees and frequencies and in the length of the data, it is somewhat difficult to quantitatively compare the measured fractional parameter changes. 
	
	The variation of the mean energy supply rate, shown in the lower right panel of Figure~\ref{fig:12} is not correlated with the level of solar activity for the investigated set of modes. This confirms the results of, e.g., \citet{Chaplin2000}, \cite{JimenezReyes2003}, and \citet{Broomhall2015}, who found that the energy supply rate to solar p modes of low harmonic degrees does not change with the level of activity. Overall there is a decrease in supply rates over the observed time period of about $8\%$. However, this decrease happened entirely before GONG month 60, hence, before the upgrade of the GONG network. Since then, it has remained at a rather constant level. A dedicated study of energy supply rates, focusing on mode sets from different frequency ranges and harmonic degree ranges, as we did in Sections~\ref{sec:3:1} and \ref{sec:3:2} for mode widths and amplitudes, will give further insight into whether the energy supply rates are truly constant over the solar cycle.

\appendix   
\section{Correction for Jumps in Mode Amplitudes}\label{app:a}
\begin{figure}
		\begin{center}
			\includegraphics[angle=90,width=0.9\textwidth]{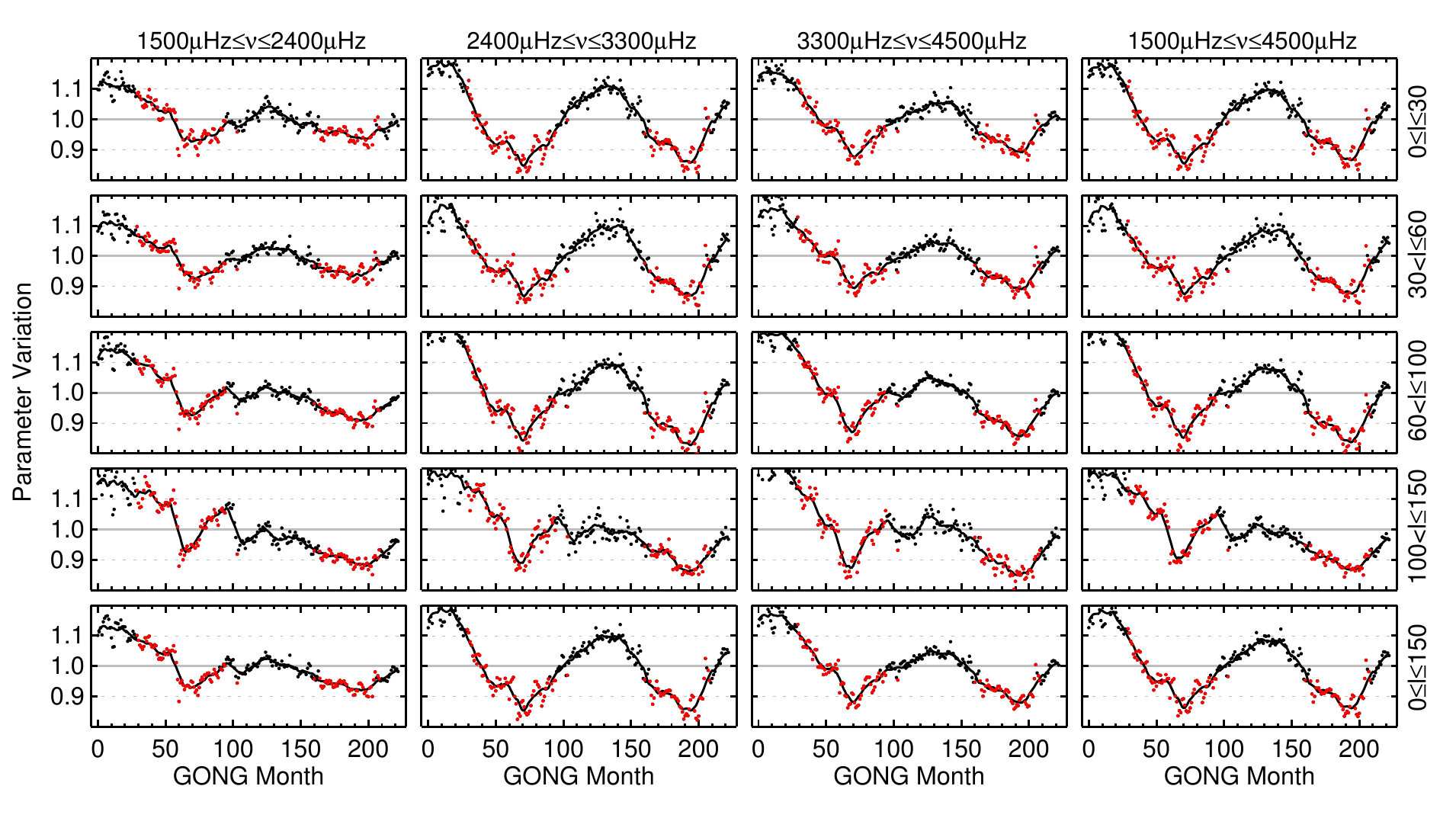}
		\end{center}
		\caption{Temporal variation of mode amplitudes which are not corrected for jumps in the data. Different ranges of harmonic degrees (rows, harmonic degree indicated to the right of the fourth column) and mode frequencies (columns, frequency range indicated above the first row) are shown. Amplitudes are normalized to the mean 	for each mode and then averaged over all modes in the respective range of frequency and degree. Months with higher than median \unit[10.7]{cm} solar radio flux are highlighted by red data points. The one year average is shown by the black solid curve. Levels of 1.1 and 0.9 of the mean are indicated by gray dashed lines. }
		\label{app:fig:1}
\end{figure}

The cameras of the GONG network were upgraded from $256\times 256$ rectangular pixels to $1024\times 1024$ square pixels around GONG month 60\footnote{For a table with the exact dates of the GONG months see \url{gong.nso.edu/data/DMAC_documentation/gongmonths.html}} \citep{Harvey1998}. Care was taken to minimize the effect of the transition between the two sets of instrumentation on the measured mode parameters by updating the data reduction and peak finding pipeline \citep{Toner2003,Hughes2016}. However, a clear jump in the mode amplitudes is visible at the time of the network upgrade. This can be seen in Figure~\ref{app:fig:1}, which shows the fractional variation of mean uncorrected mode amplitudes as a function of time for modes of different ranges of harmonic degree and mode frequency. As a complete reprocessing of the original data is not feasible, we removed this jump for each mode by matching the parameters on either side of the jump using a correction factor $C$ which is given in Equation~\ref{correction:factors:amp}. This factor is dependent on the frequency and the harmonic degree of the mode. 

There is another jump at month 100, which only affects mode amplitudes and parameters which involve mode amplitudes (see Figure~\ref{app:fig:1}). We were not able to relate this jump to a change in the instrumentation or the processing pipeline. Therefore, we patched this jump too by multiplying the mode amplitudes after month 100 with a correction factor (Equation~(\ref{correction:factors:amp})). As can be seen in Figure~\ref{app:fig:1}, the magnitude of the jumps are dependent on mode frequency (at month 60) and on harmonic degree (both at month 60 and at month 100). We included a linear term in mode frequency and a term linear in harmonic degree to the correction factor given in Equation~(\ref{correction:factors:amp}). After applying the correction for the jump at month 60 there is strong temporal variation in mode amplitudes around this time ($\pm$ 3 months). This variation does not follow a simple step-like function and is not further corrected for. Its cause is probably the incremental switch to the upgraded cameras across the network over several months.

The correction factors the mode amplitudes were multiplied by in the indicated time frame are   
\begin{align}\label{correction:factors:amp}
C &= 1 - \frac{0.03 \cdot l}{150} - \frac{0.06\cdot \nu}{\unit[4500]{\mu Hz}} -0.06 	&\text{(months 1-58)},\notag\\
C &= 1 - \frac{0.02 \cdot l}{150} - \frac{0.04\cdot \nu}{\unit[4500]{\mu Hz}} -0.04	&\text{(month 59)},\notag\\
C &= 1 - \frac{0.01 \cdot l}{150} - \frac{0.02\cdot \nu}{\unit[4500]{\mu Hz}} -0.02	&\text{(month 60)},\notag\\
C &= 1	&\text{(months 61-98)},\notag\\
C &= 1 + \frac{0.0733 \cdot l}{150} &\text{(month  100)},\notag\\
C &= 1 + \frac{0.1466 \cdot l}{150} &\text{(month 101)},\notag\\
C &= 1 + \frac{0.22 \cdot l}{150}   &\text{(months 102-223)},
\end{align}
where $l$ is the harmonic degree of the mode and $\nu$ is the frequency of the mode. Both corrections around month 60 and month 100 were found empirically. Technicians and software experts of GONG were not able to find a documented change of the instrumentation or the processing pipeline which could affect the data around month 100 in this way. As the hardware was definitely not changed at this time, the jump is probably due to an undocumented and/or unwanted change in the data processing, possibly due to an update of a software library. As the magnitude of the jump is clearly dependent on harmonic degree, see Figure~\ref{app:fig:1}, one possibility is that the spherical harmonic functions, which are used to calculate the spherical harmonic time series, changed in an unnoticed way.

No correction factor was applied to the mode widths. There is no clear jump visible around months 60 or 100 for mode widths. As can be seen in the fourth panel of the first column of Figure~\ref{fig:6}, which shows the fractional change of widths of modes with frequencies $\unit[1500$--$2400]{\mu Hz}$ and harmonic degrees 100--150, around month 45 there is a rapid change in mode widths. As this is only seen for mode widths in this small sub-set of modes, we chose not to apply a correction in this case.

\section{Results for Normalized Mode Parameter Variation}\label{app:b}

\begin{sidewaystable}
	\caption{Results from the normalized and averaged variation of mode width for different parameter ranges shown in Figure~\ref{fig:6}.\textit{Column 1--4: }Parameters ranges of the individual panels. \textit{Column 5:} Number of modes included. \textit{Column 6--7:} Extrema of the one year smoothed parameter variation \textit{Column 8:} The typical error on the individual unsmoothed data point. \textit{Column 9--10:} Spearman's correlation coefficient between independent data points of the unsmoothed parameter variation and the $F_{10.7}$ index and the associated p value. }
	\label{a2:table:1}
	\centering
		\begin{tabular}{@{\extracolsep{4pt}}cccccccccc@{}}
			\hline
			\multicolumn{2}{c}{Frequency range [\unit{$\mu$Hz}]}&\multicolumn{2}{c}{Harmonic degrees}&Number of &\multicolumn{2}{c}{Normalized width [\%]}&Mean error &&\\\cline{1-2}\cline{3-4}\cline{6-7}
			$\nu_{\text{min}}$ & $\nu_{\text{max}}$ & $l_{\text{min}}$ & $l_{\text{max}}$ & modes&   min   &   max    & [\%] & $\rho$ &           p            \\ \hline
			       1500        &        2400        &        0         &        30        &   119   & 96.7$\pm$0.2 & 105.1$\pm$0.2 &       0.6       &  0.38  &   $6.7\cdot10^{-4}$    \\
			       2400        &        3300        &        0         &        30        &   167   & 93.0$\pm$0.2 & 107.8$\pm$0.2 &       0.5       &  0.84  &      $<10^{-10}$       \\
			       3300        &        4500        &        0         &        30        &   101   & 97.4$\pm$0.2 & 103.5$\pm$0.2 &       0.5       &  0.70  &      $<10^{-10}$       \\
			       1500        &        4500        &        0         &        30        &   387   & 95.7$\pm$0.1 & 105.1$\pm$0.1 &       0.3       &  0.80  & $<10^{-10}$ 		   \\ \hline
			       1500        &        2400        &        31        &        60        &   122   & 95.0$\pm$0.1 & 106.7$\pm$0.2 &       0.4       &  0.31  &   $6.3\cdot10^{-3}$    \\
			       2400        &        3300        &        31        &        60        &   152   & 86.7$\pm$0.1 & 113.3$\pm$0.2 &       0.4       &  0.50  &   $5.2\cdot10^{-6}$    \\
			       3300        &        4500        &        31        &        60        &   123   & 95.8$\pm$0.2 & 104.0$\pm$0.2 &       0.4       &  0.50  &    $5\cdot10^{-6}$     \\
			       1500        &        4500        &        31        &        60        &   397   & 92.2$\pm$0.1 & 108.4$\pm$0.1 &       0.3       &  0.46  &   $3.6\cdot10^{-5}$    \\ \hline
			       1500        &        2400        &        61        &       100        &   130   & 96.6$\pm$0.1 & 104.7$\pm$0.1 &       0.3       &  0.32  &   $4.7\cdot10^{-3}$    \\
			       2400        &        3300        &        61        &       100        &   136   & 90.7$\pm$0.1 & 109.0$\pm$0.1 &       0.3       &  0.77  &      $<10^{-10}$       \\
			       3300        &        4500        &        61        &       100        &   88    & 94.9$\pm$0.1 & 104.9$\pm$0.2 &       0.4       &  0.80  &      $<10^{-10}$       \\
			       1500        &        4500        &        61        &       100        &   354   & 93.9$\pm$0.1 & 106.2$\pm$0.1 &       0.2       &  0.69  &      $<10^{-10}$       \\ \hline
			       1500        &        2400        &       101        &       150        &   124   & 97.6$\pm$0.1 & 103.1$\pm$0.1 &       0.3       &  0.39  &   $5.3\cdot10^{-4}$    \\
			       2400        &        3300        &       101        &       150        &    7    & 95.6$\pm$0.5 & 104.2$\pm$0.5 &       1.5       &  0.49  &   $6.4\cdot10^{-6}$    \\
			       3300        &        4500        &       101        &       150        &    6    & 95.4$\pm$0.3 & 105.4$\pm$0.6 &       1.3       &  0.78  &      $<10^{-10}$       \\
			       1500        &        4500        &       101        &       150        &   137   & 97.4$\pm$0.1 & 103.2$\pm$0.1 &       0.3       &  0.45  &   $4.3\cdot10^{-5}$    \\ \hline
			       1500        &        2400        &        0         &       150        &   495   & 96.7$\pm$0.1 & 104.3$\pm$0.1 &       0.2       &  0.36  &   $1.7\cdot10^{-3}$    \\
			       2400        &        3300        &        0         &       150        &   462   & 90.3$\pm$0.1 & 109.2$\pm$0.1 &       0.3       &  0.68  &      $<10^{-10}$       \\
			       3300        &        4500        &        0         &       150        &   318   & 96.2$\pm$0.1 & 103.6$\pm$0.1 &       0.3       &  0.74  &      $<10^{-10}$       \\
			       1500        &        4500        &        0         &       150        &  1275   & 94.4$\pm$0.1 & 105.9$\pm$0.1 &       0.2       &  0.62  &    $2\cdot10^{-9}$     \\ \hline
		\end{tabular} 
\end{sidewaystable}

\newpage

\begin{sidewaystable}
		\caption{Results from the normalized and averaged variation of mode amplitudes for different parameter ranges shown in Figure~\ref{fig:7}. \textit{Column 1--4: }Parameters ranges of the individual panels. \textit{Column 5:} Number of modes included. \textit{Column 6--7:} Extrema of the one year smoothed parameter variation. \textit{Column 8:}  The typical error on the individual unsmoothed data point. \textit{Column 9--10:} Spearman's correlation coefficient between independent data points of the unsmoothed parameter variation and the $F_{10.7}$ index and the associated p value. }
		\centering
		\label{a2:table:2}
		\begin{tabular}{@{\extracolsep{4pt}}cccccccccc@{}}
			\hline
			\multicolumn{2}{c}{Frequency range [\unit{$\mu$Hz}]}&\multicolumn{2}{c}{Harmonic degrees}&Number of &\multicolumn{2}{c}{Normalized amplitude [\%]}&Mean error&&\\\cline{1-2}\cline{3-4}\cline{6-7}
			$\nu_{\text{min}}$ & $\nu_{\text{max}}$ & $l_{\text{min}}$ & $l_{\text{max}}$ & modes&   min   &   max    &  [\%] & $\rho$ &           p            \\ \hline
			       1500        &        2400        &        0         &        30        &   119   & 94.6$\pm$0.3 & 106.2$\pm$0.4 &       1.2       & -0.82  & $<10^{-10}$ \\
			       2400        &        3300        &        0         &        30        &   164   & 87.7$\pm$0.2 & 112.5$\pm$0.3 &       0.9       & -0.94  & $<10^{-10}$ \\
			       3300        &        4500        &        0         &        30        &   101   & 93.5$\pm$0.2 & 105.1$\pm$0.3 &       0.8       & -0.88  & $<10^{-10}$ \\
			       1500        &        4500        &        0         &        30        &   384   & 91.4$\pm$0.2 & 108.5$\pm$0.2 &       0.6       & -0.92  & $<10^{-10}$ \\ \hline
			       1500        &        2400        &        31        &        60        &   122   & 94.4$\pm$0.2 & 106.9$\pm$0.2 &       0.7       & -0.78  & $<10^{-10}$ \\
			       2400        &        3300        &        31        &        60        &   152   & 88.8$\pm$0.1 & 112.5$\pm$0.2 &       0.5       & -0.88  & $<10^{-10}$ \\
			       3300        &        4500        &        31        &        60        &   123   & 93.2$\pm$0.1 & 107.1$\pm$0.2 &       0.4       & -0.88  & $<10^{-10}$ \\
			       1500        &        4500        &        31        &        60        &   397   & 91.9$\pm$0.1 & 109.0$\pm$0.1 &       0.3       & -0.87  & $<10^{-10}$ \\ \hline
			       1500        &        2400        &        61        &       100        &   130   & 94.6$\pm$0.1 & 107.5$\pm$0.2 &       0.5       & -0.76  & $<10^{-10}$ \\
			       2400        &        3300        &        61        &       100        &   136   & 86.7$\pm$0.1 & 114.1$\pm$0.2 &       0.5       & -0.91  & $<10^{-10}$ \\
			       3300        &        4500        &        61        &       100        &   88    & 90.3$\pm$0.1 & 109.9$\pm$0.2 &       0.5       & -0.89  & $<10^{-10}$ \\
			       1500        &        4500        &        61        &       100        &   354   & 90.6$\pm$0.1 & 110.6$\pm$0.1 &       0.3       & -0.90  & $<10^{-10}$ \\ \hline
			       1500        &        2400        &       101        &       150        &   124   & 94.0$\pm$0.1 & 107.4$\pm$0.2 &       0.5       & -0.66  & $<10^{-9}$  \\
			       2400        &        3300        &       101        &       150        &    7    & 90.5$\pm$0.4 & 108.9$\pm$0.7 &       1.8       & -0.82  & $<10^{-10}$ \\
			       3300        &        4500        &       101        &       150        &    6    & 89.1$\pm$0.4 & 112.2$\pm$0.7 &       1.7       & -0.87  & $<10^{-10}$ \\
			       1500        &        4500        &       101        &       150        &   137   & 93.8$\pm$0.1 & 107.7$\pm$0.2 &       0.5       & -0.70  & $<10^{-10}$ \\ \hline
			       1500        &        2400        &        0         &       150        &   495   & 95.0$\pm$0.1 & 106.9$\pm$0.1 &       0.4       & -0.81  & $<10^{-10}$ \\
			       2400        &        3300        &        0         &       150        &   459   & 87.8$\pm$0.1 & 112.9$\pm$0.1 &       0.4       & -0.92  & $<10^{-10}$ \\
			       3300        &        4500        &        0         &       150        &   318   & 92.4$\pm$0.1 & 107.3$\pm$0.1 &       0.3       & -0.90  & $<10^{-10}$ \\
			       1500        &        4500        &        0         &       150        &  1272   & 91.8$\pm$0.1 & 109.1$\pm$0.1 &       0.2       & -0.91  & $<10^{-10}$ \\ \hline
		\end{tabular}
\end{sidewaystable}

\clearpage

\section{Results of fits to mode amplitudes for different ranges of harmonic degrees}\label{app:c}
To investigate the degree dependence of the mode amplitudes, we separated the modes into groups of typically 10 harmonic degrees, see the first column of Table~\ref{a3:table:1}. We excluded modes with $l=0,1$ because their amplitudes are significantly different from those of modes with $2\le l \le10$ (see Fig. 8 of \citealp{Howe2003} for mean mode heights of low-degree modes from GONG). In order to get a good fit of the mode amplitudes for modes with $l>110$, we grouped together the range covering degrees $111 \le l \le 150$. We included all modes which are present in at least 50\% of the GONG months. Lowering the presence rate from 100\% was necessary because there were not enough modes present in certain frequency ranges and the fits did not converge properly. We describe the Voigt profile and the fit results of the complete set of modes $l=2-150$ in Section~\ref{sec:3:3}. There, we also present a fit to the amplitudes of the group $31\le l\le 40$ (bottom panel of Figure~\ref{fig:8}).

\begin{figure}
	\begin{center}
		\includegraphics[width=\textwidth]{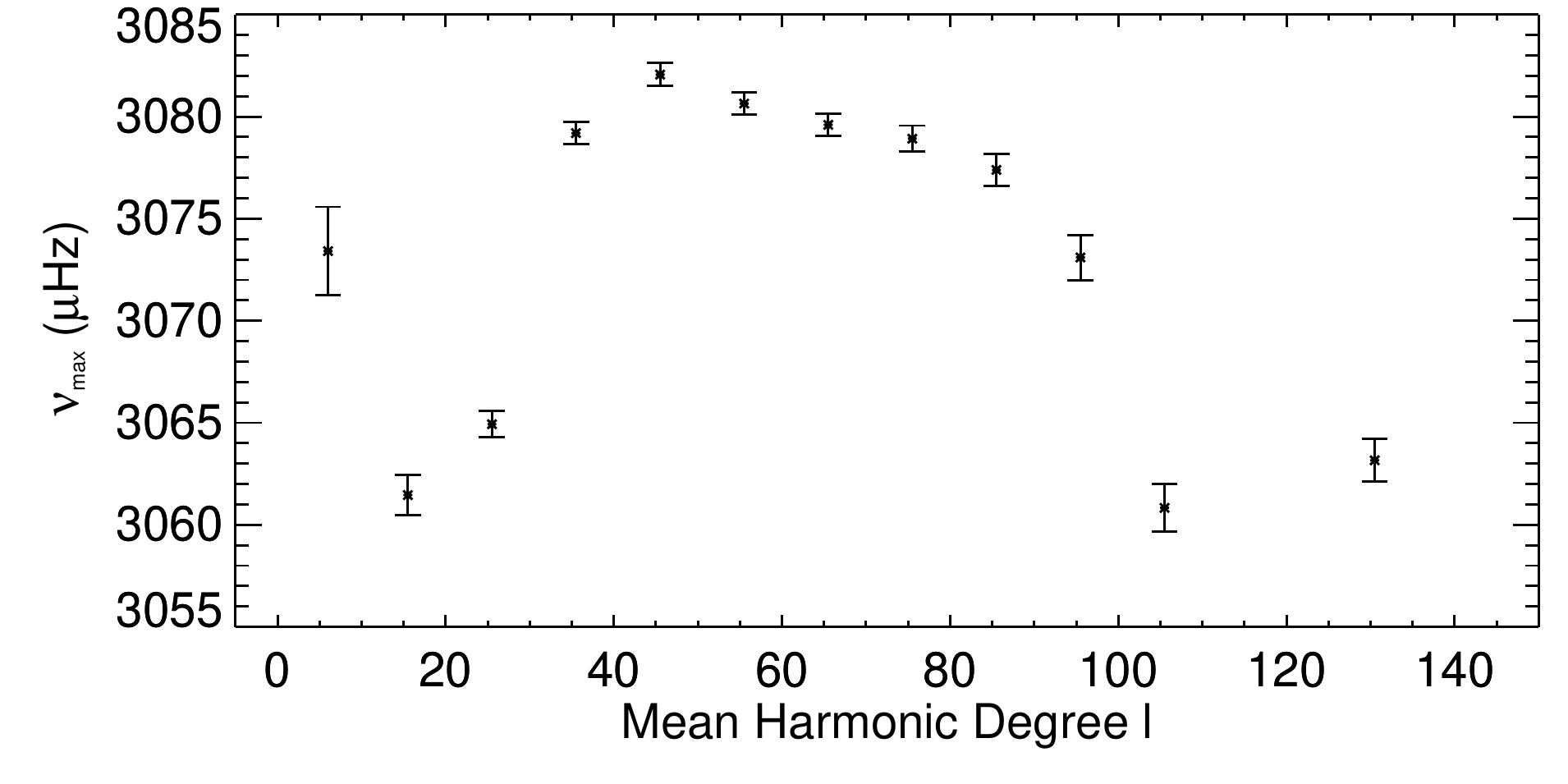}
	\end{center}
	\caption{Frequency of maximum amplitude $\nu_{\text{max}}$ for different ranges of harmonic degrees. }
	\label{app:fig:2}
\end{figure}

\begin{figure}
	\begin{center}
		\includegraphics[width=\textwidth]{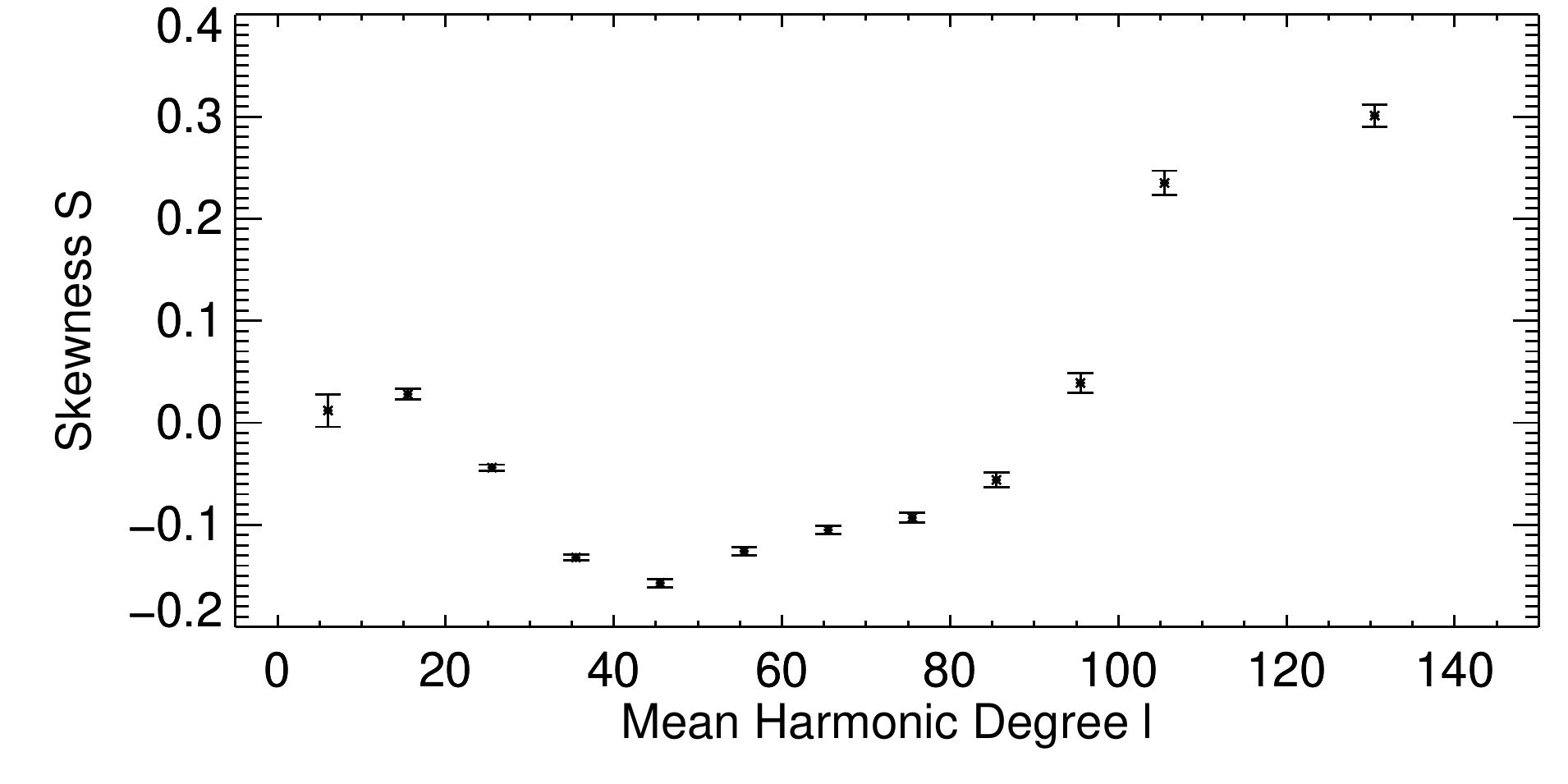}
	\end{center}
	\caption{Same as Figure~\ref{app:fig:2} but for the skewness $S$.}
	\label{app:fig:5}
\end{figure}

Figures~\ref{app:fig:2}-\ref{app:fig:5} show the fit parameters frequency of maximum amplitude $\nu_{\text{max}}$, skewness $S$, width of the asymmetric Voigt $\Sigma$, and amplitude parameter $a$ as functions of the mean harmonic degree of the ranges listed in the first column of Table~\ref{a3:table:1}. 

In Figure~\ref{app:fig:2} it can be seen that $\nu_{\text{max}}$ first decreases from the group of lowest degree modes to second and third group before it increases. It is then more or less constant for modes with $31 \le l \le 90$ and decreases again for higher degree modes. The skewness $S$, shown in Figure~\ref{app:fig:5}, is close to 0 for modes $l<21$, negative for modes with $21 \le l \le 90$, and turns positive for modes with $l>90$. The frequency of maximum amplitude $\nu_{\text{max}}$ and the skewness $S$ are strongly anti-correlated with a rank correlation of $\rho=-0.91$ and a p value $<3\cdot10^{-5}$.

The width of the Voigt profile changes by only about 3\% for modes with $2\le l \le 110$. The much larger $\Sigma$ of the group with $111\le l 150$ is probably due to the wider range of modes included in this group (see also Section~\ref{sec:3:3}). The amplitude parameter $a$ (Figure~\ref{app:fig:4}) increases for the first three groups of harmonic degrees from $\unit[2800\pm14]{m^2\,s^{-2}\,Hz^{-1}}$ for $2\le l \le 10$ to $\unit[3424\pm4]{m^2\,s^{-2}\,Hz^{-1}}$ for $31\le l \le 40$. Compared to this large increase, the amplitudes parameter changes very little for the groups of modes of higher degrees. 


\begin{figure}
	\begin{center}
		\includegraphics[width=\textwidth]{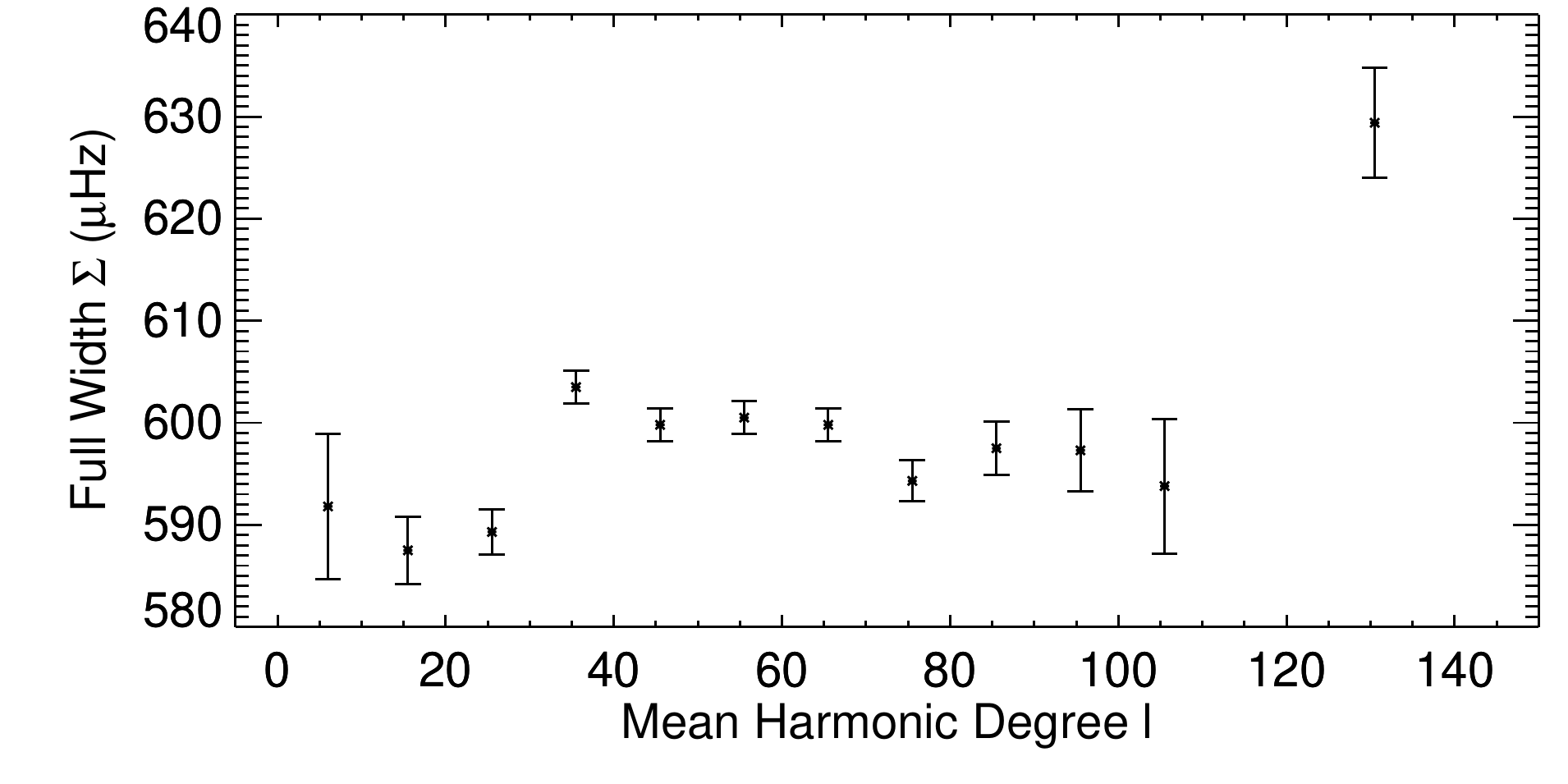}
	\end{center}
	\caption{Same as Figure~\ref{app:fig:2} but for the width of the Voigt profile $\Sigma$.}
	\label{app:fig:3}
\end{figure}
\begin{figure}
	\begin{center}
		\includegraphics[width=\textwidth]{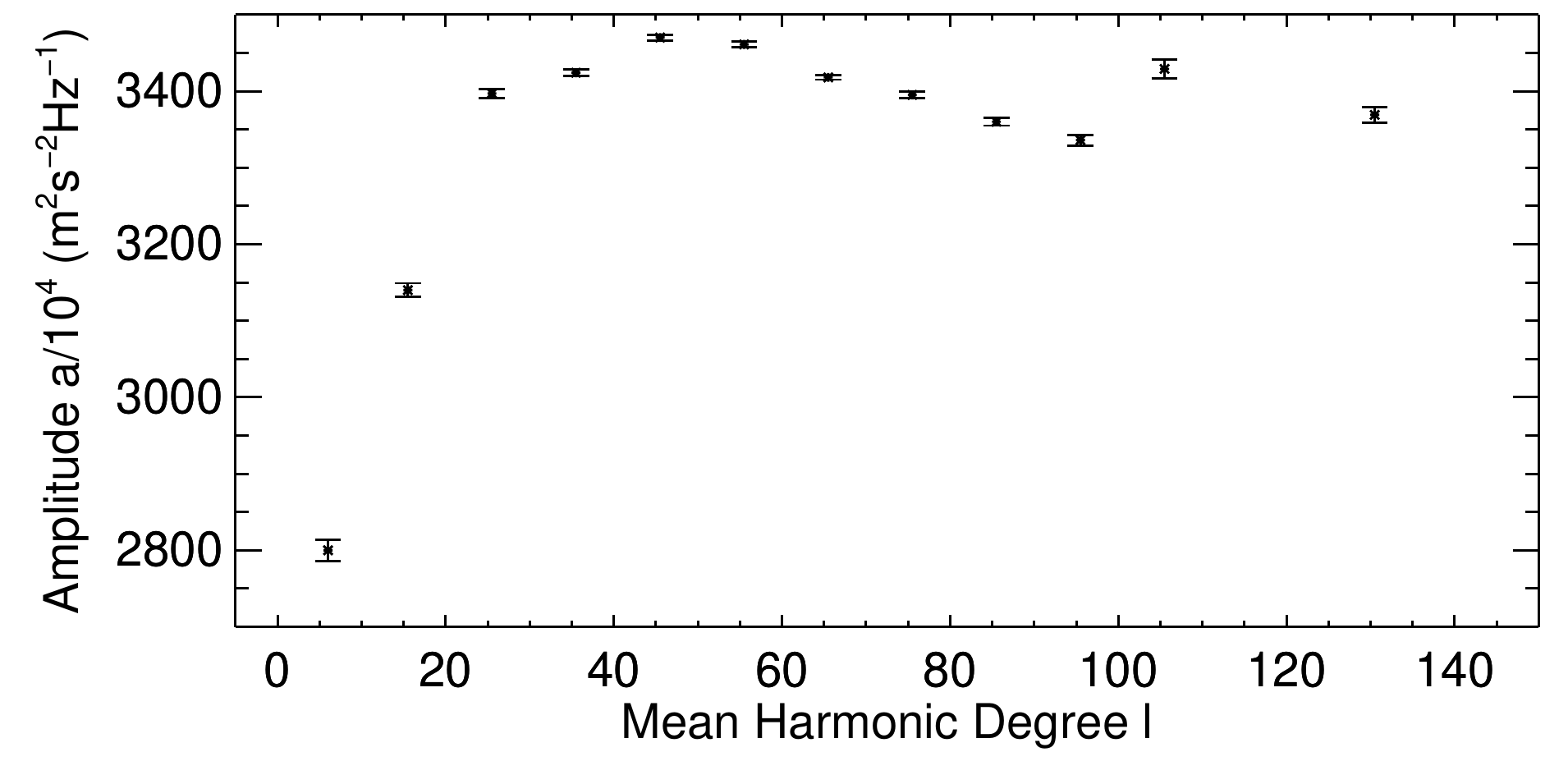}
	\end{center}
	\caption{Same as Figure~\ref{app:fig:2} but for the amplitude parameter $a$.}
	\label{app:fig:4}
\end{figure}

\clearpage

\begin{sidewaystable}
	\caption{Parameters of fits to the frequency distribution of mode amplitudes for different ranges of harmonic degrees.}\label{a3:table:1}
	\begin{tabular}{cccccccccc}
		\hline 
		Range of &$\nu_{\text{max}}$ & $\sigma$&$\gamma$ & $\Sigma$ &$a/{10^4}$& $b$ & $S$ &Number&$\chi^2_{\text{red}}$\\ 
		degrees $l$ &\text{[}$\mu$Hz]& [$\mu$Hz]&[$\mu$Hz]&[$\mu$Hz] & \text{[}$\unit{m^2\,s^{-2}\,Hz^{-1}}$] & [$\unit{m^2\,s^{-2}\,Hz^{-1}}$]  & &of modes&\\
		\hline 
		2--10 & 3073.41$\,\pm\,$2.17 & 166.5$\,\pm\,$2.9 & 161.0$\,\pm\,$2.6 & 591.8$\,\pm\,$ 7.1   & 2800$\,\pm\,$ 14 &  -640$\,\pm\,$ 19 &  0.012$\,\pm\,$0.016 & 132 & 9.3\\
		11--20 & 3061.46$\,\pm\,$0.98 & 169.2$\,\pm\,$1.4 & 153.5$\,\pm\,$1.1 & 587.5$\,\pm\,$ 3.3  & 3140$\,\pm\,$  9 &  -611$\,\pm\,$  7 &  0.028$\,\pm\,$0.005 & 175 & 6.4 \\
		21--30 & 3064.93$\,\pm\,$0.65 & 168.5$\,\pm\,$0.9 & 156.1$\,\pm\,$0.7 & 589.3$\,\pm\,$ 2.2  & 3397$\,\pm\,$  6 &  -655$\,\pm\,$  5 & -0.044$\,\pm\,$0.003 & 172 &10.4 \\
		31--40 & 3079.19$\,\pm\,$0.54 & 177.0$\,\pm\,$0.7 & 152.6$\,\pm\,$0.5 & 603.5$\,\pm\,$ 1.6  & 3424$\,\pm\,$  4 &  -630$\,\pm\,$  3 & -0.132$\,\pm\,$0.003 & 157 &10.8 \\\hline
		41--50 & 3082.06$\,\pm\,$0.56 & 174.9$\,\pm\,$0.7 & 153.3$\,\pm\,$0.5 & 599.8$\,\pm\,$ 1.6  & 3470$\,\pm\,$  4 &  -638$\,\pm\,$  3 & -0.157$\,\pm\,$0.004 & 134 & 9.9 \\
		51--60 & 3080.64$\,\pm\,$0.55 & 175.4$\,\pm\,$0.7 & 153.0$\,\pm\,$0.5 & 600.5$\,\pm\,$ 1.6  & 3461$\,\pm\,$  4 &  -619$\,\pm\,$  3 & -0.126$\,\pm\,$0.004 & 118 &13.1 \\
		61--70 & 3079.60$\,\pm\,$0.54 & 174.4$\,\pm\,$0.7 & 154.1$\,\pm\,$0.5 & 599.8$\,\pm\,$ 1.6  & 3418$\,\pm\,$  3 &  -594$\,\pm\,$  3 & -0.105$\,\pm\,$0.004 & 113 & 9.7 \\
		71--80 & 3078.92$\,\pm\,$0.64 & 165.6$\,\pm\,$0.8 & 164.1$\,\pm\,$0.7 & 594.3$\,\pm\,$ 2.0  & 3395$\,\pm\,$  4 &  -660$\,\pm\,$  4 & -0.093$\,\pm\,$0.005 &  99 & 9.0 \\\hline
		81--90 & 3077.39$\,\pm\,$0.78 & 160.9$\,\pm\,$1.1 & 173.6$\,\pm\,$0.9 & 597.5$\,\pm\,$ 2.6  & 3360$\,\pm\,$  5 &  -706$\,\pm\,$  5 & -0.056$\,\pm\,$0.007 &  89 & 8.8 \\
		91--100 & 3073.10$\,\pm\,$1.11 & 150.3$\,\pm\,$1.7 & 189.0$\,\pm\,$1.5 & 597.3$\,\pm\,$ 4.0 & 3336$\,\pm\,$  7 &  -734$\,\pm\,$  8 &  0.039$\,\pm\,$0.010 &  79 &13.2  \\
		101--110 & 3060.83$\,\pm\,$1.17 & 127.4$\,\pm\,$3.0 & 217.0$\,\pm\,$2.7 & 593.8$\,\pm\,$ 6.6& 3429$\,\pm\,$ 12 &  -780$\,\pm\,$ 10 &  0.235$\,\pm\,$0.012 &  80 &11.2  \\
		111--150 & 3063.16$\,\pm\,$1.04 & 142.6$\,\pm\,$2.3 & 220.6$\,\pm\,$2.3 & 629.4$\,\pm\,$ 5.4& 3369$\,\pm\,$ 10 &  -804$\,\pm\,$  9 &  0.301$\,\pm\,$0.011 & 195 &19.1  \\ \hline
	\end{tabular}  
\end{sidewaystable} 

\clearpage
%
\begin{acks}
This work utilizes data obtained by the Global Oscillation Network Group (GONG) Program, managed by the National Solar Observatory, which is operated by AURA, Inc. under a cooperative agreement with the National Science Foundation. The data were acquired by instruments operated by the Big Bear Solar Observatory, High Altitude Observatory, Learmonth Solar Observatory, Udaipur Solar Observatory, Instituto de Astrof\'isica de Canarias, and Cerro Tololo Interamerican Observatory. RK \& AMB acknowledge the support of the STFC consolidated grant ST/P000320/1. The research leading to these results has received funding from the European Research Council (ERC) under the European Union’s Seventh Framework Program (FP/2007-2013) / ERC Grant Agreement n. 307117. This work was supported by the SOLARNET project funded by the European Commissionʼs FP7 Capacities Programme under the Grant Agreement 312495.
\end{acks}

%
%

\bibliographystyle{spr-mp-sola}
\bibliography{references}

\end{article} 
\end{document}